\documentclass[journal]{IEEEtran}
\usepackage{hyperref}
\usepackage{todonotes} 
\newcommand{\one}[1]{{\color{black}#1}}
\newcommand{\two}[1]{{\color{black}#1}}
\newcommand{\three}[1]{{\color{black}#1}}
\usepackage{cite}
\usepackage{mathtools}

\usepackage[ruled,vlined]{algorithm2e}
\usepackage{subcaption}
\usepackage{enumerate}
\usepackage[inline]{enumitem}
\makeatletter
\let\old@ps@headings\ps@headings
\let\old@ps@IEEEtitlepagestyle\ps@IEEEtitlepagestyle
\def\psccfooter#1{%
    \def\ps@headings{%
        \old@ps@headings%
        \def\@oddfoot{\strut\hfill#1\hfill\strut}%
        \def\@evenfoot{\strut\hfill#1\hfill\strut}%
    }%
    \def\ps@IEEEtitlepagestyle{%
        \old@ps@IEEEtitlepagestyle%
        \def\@oddfoot{\strut\hfill#1\hfill\strut}%
        \def\@evenfoot{\strut\hfill#1\hfill\strut}%
    }%
    \ps@headings%
}
\makeatother

\begin{document}
%
\title{Low Voltage Customer Phase Identification Methods Based on Smart Meter Data}

\author{\IEEEauthorblockN{Alexander Hoogsteyn\IEEEauthorrefmark{1},
Marta Vanin\IEEEauthorrefmark{1}\IEEEauthorrefmark{2},
Arpan Koirala\IEEEauthorrefmark{1}\IEEEauthorrefmark{2}, Dirk Van Hertem\IEEEauthorrefmark{1}\IEEEauthorrefmark{2}}
\IEEEauthorblockA{\IEEEauthorrefmark{1} KU Leuven, Dept. of Electrical Engineering (ESAT), Kasteelpark Arenberg 10, 3001, Heverlee, Belgium}
\IEEEauthorblockA{\IEEEauthorrefmark{2} EnergyVille, Thor Park 8310, 3600, Genk, Belgium}
}

\maketitle


\begin{abstract}
The increased deployment of distributed energy generation and the integration of new, large electric loads such as electric vehicles and heat pumps challenge the correct and reliable operation of low voltage distribution systems. To tackle potential problems, active management solutions are proposed in the literature, which require distribution system models that include the phase connectivity of all the consumers in the network. However, information on the phase connectivity is in practice often unavailable. In this work, several voltage and power measurement-based phase identification methods from the literature are implemented. A consistent comparison of the methods is made across different smart meter accuracy classes and smart meter penetration levels using publicly available data. Furthermore, a novel method is proposed that makes use of ensemble learning and that can combine data from different measurement campaigns. \two{The results indicate that generally better results are obtained with voltage data compared to power data from smart meters of the same accuracy class. If power data is available too, the novel ensemble method can improve the accuracy of the phase identification obtained from voltage data alone.}

\end{abstract}

\begin{IEEEkeywords}
clustering,  ensemble learning, low voltage distribution system, phase identification, smart meter
\end{IEEEkeywords}

%
\IEEEpeerreviewmaketitle

\section{Introduction}
\label{sec:intro}
%
%
%
%
\IEEEPARstart{T}{he} shift to more distributed generation~\cite{noauthor_renewables_nodate} and integration of new large loads such as electric vehicles and heat pumps, can lead to various problems in distribution systems (DS), such as increased imbalance and overvoltages. Historically, imbalance was not considered a problem since DS were typically underutilized and characterized by modest and unidirectional power flows. The emergence of PV installations and larger loads implies more utilization and unpredictability, which might push the DS state beyond the acceptable operational limits.

To cope with the aforementioned issues, recent research on DS advocates for an improved system exploitation and control, through the so-called Active DS Management. Examples are optimal voltage control by means of tap changes \cite{long_performance_2015} or resilient restoration \cite{dubey_paving_2020}. A distribution system operator (DSO) could also preventively reconfigure the phase connection of its grid, such that a more balanced configuration is obtained. For such applications, it is necessary to identify the present phase connectivity.

Phase identification methods try to classify to which phase(s) a consumer is connected, usually with the help of field measurements. Manual determination of the phase labeling is often not viable because of the cost of this labour-intensive approach, especially since the majority of the cables are underground. Furthermore, cables are not color-coded. In literature, a wide variety of data-driven approaches are proposed. In \cite{chen_design_2011, wen_phase_2015}, field measurement-based methods are described that use communicating phasor measurement units (PMU) between the consumer and the transformer. However, these devices are quite expensive and unlikely to be widely available in DS in the near future, especially at the low voltage levels. Smart meters, which are already being rolled out, are a more realistic option, at least in the near future\cite{watson_use_2016,luan_distribution_2013,padullaparti_considerations_2019}. Developing practically applicable methods that rely on realistic assumptions on available smart meter data is still an open research topic.

\one{The remainder of the paper is structured as follows: In Section \ref{sec:litrev}, the current literature on smart meter (SM) data based phase identification methods is reviewed. In Section \ref{sec:probfor}, the mathematical formulation of the phase identification methods used in the present work is given, including our novel ensemble learning method. In Section \ref{sec:numill}, the performance of the different methods is illustrated under a broad range of scenarios. Finally, conclusions are drawn}.

\section{\one{Related work}}
\label{sec:litrev}
Phase identification methods that use SM data can be categorized into 3 groups based on whether they use: \begin{enumerate*}
    \item [a)] mixed-integer programming (MIP) approaches typically using power data,    \item [b)] machine learning, with voltage data (MLV), or
    \item [c)] machine learning, with power data (MLP).
\end{enumerate*}
In \two{\cite{arya_phase_2011, arya_inferring_2013, pappu_identifying_2018, zhou_consumer_2020, heidari-akhijahani_phase_2021}}, a MIP approach is used to solve the phase identification problem. These optimization-based implementations need measurements of the consumers' demand and the distribution transformer's supply, and use the principle of conservation of power to determine which consumer is connected to which phase. The algorithm optimizes the allocation of the set of consumers to the three phases by minimizing the difference between the total feeder demand and the transformer supply. These MIP methods are inherently inferior at handling missing data or incomplete datasets, and typically assume that all users have a smart meter and there are no electricity thefts or similar phenomena \cite{pappu_identifying_2018}. MIP-based methods that require both voltage and power data have also been proposed \cite{zhou_consumer_2020}. The authors propose  a method which allocates the phases firstly using voltage correlation, afterwards optimization techniques are used that is initiated with the voltage-based solution and then iteratively solved.

To make these methods more rigorous, line losses should be modelled. However, this requires knowledge of line impedance values, which are typically not known in sufficient detail in low voltage networks. While \cite{arya_phase_2011, zhou_consumer_2020} do not rigorously model the network physics, in \cite{heidari-akhijahani_phase_2021}, a MIP formulation is proposed which explicitly adds linearized power flow equations as constraints.

In \two{\cite{pezeshki_correlation_2012, seal_automatic_2011, luan_smart_2015, pezeshki_consumer_2012, short_advanced_2013, watson_use_2016, arya_voltage-based_2013,  mitra_voltage_2015, wang_phase_2016, ma_phase_2018, liu_practical_2020, blakely_spectral_2019, Simonovska}} MLV approaches are proposed. The algorithm in \cite{pezeshki_correlation_2012, seal_automatic_2011} uses the correlation between time-series measurements of a customer's voltage and the transformer voltage time series of each phase to distinguish the phase connection. It relies on the principle that consumers connected to the same phase will have similar voltage time series and therefore have a higher Pearson correlation coefficient. The advantage of using such machine learning methods is that they can deal with incomplete data better, since they do not use the conservation of power. The authors in \cite{luan_smart_2015} improve upon such an approach by leveraging the known topology by accounting for the voltage drop corresponding to the distance between the feeder and the consumer's connection. In \cite{pezeshki_consumer_2012}, the author uses other consumers as reference instead of transformer measurements. This avoids the need to collect transformer measurements which are often not available. Such an approach does require data of a three-phase consumer. \one{One problem MLV methods face is that voltage variations are relatively small near the transformer. Load correction is applied by \cite{short_advanced_2013} and \cite{watson_use_2016} to mitigate this problem. In load correction, power data is used in addition to the voltage data of a consumer to estimate the voltage drop on the phase the consumer is connected to.} 

References \cite{arya_voltage-based_2013, mitra_voltage_2015, wang_phase_2016} use K-means clustering, ref.~\cite{Simonovska} combines K-means clustering and principal component analysis. Ref. \cite{ma_phase_2018, liu_practical_2020} use spectral clustering (a variant of K-means) to group the voltage time-series measurements in three clusters, based on their similarity. Each cluster corresponds to one of the phases. This removes the need for a reference voltage profile entirely. \one{In \cite{blakely_spectral_2019, blakely_2020}, a sliding window ensemble learning method is used, which combines spectral clustering results using voltage data of different days.}

MLP methods were proposed in \two{\cite{hosseini_machine_2020, jayadev2016novel, xu_phase_2018}} that use concepts from ML such as correlation. Such methods can cope with missing data and thus mitigate the disadvantages of MIP methods while using power measurements. These measurements are more often available than the voltage in a real-world setting because they are needed for billing purposes. The approaches rely on the principle that a consumers' demand is more correlated to the supply of the phase it is connected to. Consequently, the methods also need measurements of the power supplied from each transformer's phase. \two{The advantages and disadvantages of the categories of the discussed methods are summarized in Table \ref{tab:method_comp}. }

\one{Besides \cite{therrien_assessment_2021},} research work that copes with the rigorous comparison of the available phase identification techniques is currently scarce. This paper aims to do such a comparison. The main contributions of this paper are: \begin{itemize}
    \item Different existing techniques for phase identification in DS using SM data are compared in a thorough and consistent manner. The comparisons are carried out over different SM accuracy classes and data availability scenarios, on multiple non-synthetic European feeders  with different characteristics.
    \item A novel algorithm for SM based phase-identification is proposed that makes use of ensemble learning to combine results from voltage and power-based methods. The ensemble method improves upon existing power-based techniques and optionally takes voltage data into account to yield further improvements. \one{To the authors' knowledge, it is the first ensemble learning algorithm that combines results from voltage-based and power-based methods.}
\end{itemize}

\begin{table}[t]
    \caption{Advantages and disadvantages of phase identification methods.}
    \begin{tabular}{p{0.05\textwidth}p{.2\textwidth}p{0.2\textwidth}}
    \hline
        & \textbf{Advantages} & \textbf{Disadvantages} \\
        \hline
        \textbf{MIP} \cite{arya_phase_2011,arya_inferring_2013, pappu_identifying_2018, zhou_consumer_2020, heidari-akhijahani_phase_2021} & \begin{itemize}
           \item Can make use of existing optimization techniques
           \item Load data more often available
        \end{itemize}        & \begin{itemize}
            \item Hard to cope with missing data
            \item Need to model the losses to be accurate
        \end{itemize} \\
        \textbf{MLV} \cite{pezeshki_correlation_2012, seal_automatic_2011, luan_smart_2015, pezeshki_consumer_2012, short_advanced_2013, watson_use_2016, arya_voltage-based_2013,  mitra_voltage_2015, wang_phase_2016, ma_phase_2018, liu_practical_2020, blakely_spectral_2019}& \begin{itemize}
            \item  Can cope with missing data
            \item Can make use of existing machine-learning techniques
            
        \end{itemize} & \begin{itemize}
            \item Voltage data often not available
            \item Decreased performance for short feeders and close to the transformer
        \end{itemize} \\
        \textbf{MLP} \cite{xu_phase_2018,hosseini_machine_2020, jayadev2016novel}& \begin{itemize}
            \item Can cope with missing data
            \item Load data more often available
        \end{itemize} & \begin{itemize}
            \item Needs novel algorithms 
            \item Lower performance on feeders with a high amount of connections
        \end{itemize}\\
        \hline
    \end{tabular}
    
    \label{tab:method_comp}
\end{table}

\section{Problem formulation}
\label{sec:probfor}
Let $i$ be each of the $N$ customers in the radial low voltage distribution network, and $j$ be one of the three phases: $j\in\{a,b,c\}$. Customer voltage measurements $U$ are collected for the MLV method, and consumer $H$ and transformer $P$ power measurements for the MLP method. In the proposed ensemble method, all of these can be used. All measurements are averages over a period of time that depends on the meter resolution. Let $h_{t,i}$ and $u_{i,t}$ represent the average power and voltage measurement of customer $i$ for the time period $t \in \{1, \ldots, T\}$. Finally, let $p_{t,j}$ be the transformer power measurement for phase $j$, time $t$, such that:
\begin{equation}
 H=\left[\begin{array}{l l l l}
 h_{1,1} & \cdots & h_{1,N}  \\
 \vdots & \ddots & \vdots \\
 h_{T,1} & \cdots & h_{T,N} \\
\end{array}\right]
\label{eq:H}
\end{equation}
\begin{equation}
P=\left[\begin{array}{l l l l}
p_{1,a} & p_{1,b} & p_{1,c} \\
\vdots & \vdots & \vdots \\
p_{T,a} & p_{T,b} & p_{T,c} \\
\end{array}\right]
\label{eq:P}
\end{equation}
\begin{equation}
U=\left[\begin{array}{l l l l}
u_{1,1} & \cdots & u_{1,N}\\
 \vdots & \ddots & \vdots \\
u_{T,1} & \cdots & u_{T,N}\\
\end{array}\right]
\label{eq:U}
\end{equation}
Gaussian noise $E \sim \mathcal{N}(0,\,\sigma_{error}^{2})$ is added to $U$, $H$ and $P$ to simulate the accuracy classes of SM specified by the International Electrotechnical Commission (IEC). \two{50 Monte Carlo runs are performed with different noise samples, and the average obtained accuracies are reported.} The accuracy classes $0.1$, $0.2$, $0.5$, and $1.0$ state that the maximal measurement error ($\delta_{class}$) needs to be below $\pm0.1$, $\pm0.2$, $\pm0.5$, or $\pm1.0$ \% of the nominal value of the meter (indicated as $U_{n}$ or $P_n$). To mimic the requirement set by the IEC, which specifies a maximal measurement error, we assume that the maximal error corresponds to 3 sigma of the chosen distribution, as such:
\begin{equation}
    \sigma_{error,U} =\frac{1}{3} \delta_{class}U_{n}
    \label{eq:sdev}
\end{equation}
\begin{equation}
    \sigma_{error,P} =\frac{1}{3} \delta_{class}{P_{n}}
    \label{eq:sdev_p}
\end{equation}

Two additional high accuracy classes, 0.2s and 0.5s, are considered. These classes are especially more accurate in low loading conditions since the maximum deviation of a measured value is defined as a fraction of the actual loading and not as a fraction of the nominal rating (between 20\% and 100\% of the nominal rating). This is useful for MLP methods, since power demand is rarely close to the nominal value, unlike voltages. For these classes, the standard deviation of the artificially added error is found as follows:
\begin{equation}
    \sigma_{error,P} = \begin{cases}
    \frac{1}{3} \delta_{class}\frac{p_{i,t}}{P_{base}} & p_{i,t} > 0.2 P_{n} \\
    \frac{1}{3} \delta_{class}\frac{ 0.2 P_{n}}{P_{base}} &  p_{i,t} \leq 0.2 P_{n} \\
    \end{cases}
    \forall i,t.
    \label{eq:sdev_power}
\end{equation}

\two{It is assumed that the user phase connectivity does not change over the measurement collection time. This is a realistic assumption, as reconfiguration actions are not performed often in low voltage networks, e.g., due to the lack of automatic switching devices. System operators can reasonably schedule their operations such that the two activities do not overlap in time.}

\subsection{Voltage-based methods (MLV)}
The first MLV method implemented is that of \cite{pezeshki_correlation_2012} (see Algorithm 1). The method calculates the Pearson correlation \eqref{eq:pearson} between the voltage time series $U_{i}= [u_{1,i}, u_{2,i}, \ldots]$ of each consumer $i$ and a reference voltage time series $U_{ref,j}$, which can be from a three-phase reference customer or a transformer measurement. With $\mu(U_{i})$ the average voltage of $i$ and $\sigma(U_{i})$ its standard deviation. Similarly, $\mu(U_{ref,j})$ the average voltage of reference phase $j$ and $\sigma(U_{ref,j})$ its standard deviation. Users are assigned to the reference phases with which they have the highest correlation.
\begin{equation}
\begin{aligned}
    \rho(U_{i},U_{ref,j}) &= \frac{(U_{i}-\mu(U_{i}))( U_{ref,j}-\mu(U_{ref,j}))}{\sigma(U_{i}) \sigma(U_{ref,j})} \\
    &= \frac{\sum^T_{t=1} (u_{t,i}-\mu(U_{i}))(u_{t,ref,j}-\mu(U_{ref,j}))}{\sigma(U_{i}) \sigma(U_{ref,j})}
    \label{eq:pearson}
\end{aligned}
\end{equation}

\begin{algorithm}[htb]
\SetAlgoLined
\DontPrintSemicolon
\KwData{$U_{i}=[u_{1,i},\ldots,u_{T,i}]$ for all consumers $i$, \;
\indent $U_{ref,j}=[u_{ref,1,j},\ldots,u_{ref,T,j}]$ for each phase $j$} \;
 \For{Each consumer $i \in \{1,\ldots,N \}$}{
    \For{Each phase $ j \in \{a,b,c\}$}{
    Calculate $\rho(U_{i},U_{ref,j})$ using (\ref{eq:pearson}) \;
    }
 Allocate $i$ with highest  $\rho(U_{i},U_{ref,j})$ to phase $j$  \;
 }
\label{alg:kmeans}
\caption{Pearson correlation \cite{pezeshki_correlation_2012}.}
\end{algorithm}
Another MLV approach to voltage-based phase identification is K-means clustering \cite{arya_voltage-based_2013}. This aims to partition a data set in $k$ clusters by allocating voltage time-series $U_{i}$ to a `mean profile', also called cluster center $C_{k}$. In this case there are 3 clusters, one for each phase. The Euclidean distance between each data point and its cluster center is minimized (\ref{eq:kmeans}). This is obtained using an iterative process until the stopping condition is met set by $\epsilon$, as described in Algorithm 2. Alternative time-shift invariant distance metrics have been explored, such as dynamic time warping. Such metrics can lead to better time-series clustering results~\cite{kate_dynamic_2015}, but this was not the case in our examined numerical illustrations. \two{The choice to implement K-means method is based on literature considerations, as K-means is by far and large the most widely employed clustering method for phase identification. Other clustering approaches have been investigated, such as hierarchical clustering and Gaussian mixture model, but their results are not shown as they present no valuable difference to those of K-means. The implementation of all methods explored in this work is available open source\footnote{\url{https://github.com/AlexanderHoogsteyn/PhaseIdentification}}.} 
\begin{equation}
\underset{\ell_{1}, \ldots, \ell_{k}}{\min } \sum_{j=1}^{k} \sum_{\mathbf{u}_{i} \in \ell_{j}} \lVert \mathbf{U}_{i}-\mathbf{C}_{j}\rVert^{2}_{2}
\label{eq:kmeans}
\end{equation}
\begin{algorithm}[htb]
\SetAlgoLined
\DontPrintSemicolon
\KwData{$U_{i}=[u_{1,i},\ldots,u_{T,i}]$ for all consumers $i$} \;
Randomly initialize cluster centers $C_{j}$\;
 \While{$\sum_{j=1}^{k} \sum_{\mathbf{U}_{i} \in \ell_{j}}||\mathbf{U}_{i}-\mathbf{C}_{j}||^{2}_{2} > \epsilon$}{
 \For{Each consumer $i \in \{1,\ldots,N \}$}{
    \For{Each cluster $ k \in \{k_{1},k_{2},k_{3}\}$}{
    Calculate distance to cluster center $d_{ik}  \leftarrow ||U_{i}-C_{k}||_{2}$ \;
    }
 Allocate consumer $i$ to $\ell_{k}$ with smallest $d_{ik}$ \;
 }
 \For{Each cluster $k \in \{k_{1},k_{2},k_{3}\}$}{
 Calculate mean $M_{k}  \leftarrow \frac{1}{|\ell_{k}|}\sum_{\mathbf{U}_{i} \in
 \ell_{k}}\mathbf{U}_{i}$ \;
 Set mean as new cluster center $C_{k}  \leftarrow M_{k}$ \;
 }
 }
\label{alg:kmeans}
\caption{K-means clustering \cite{arya_voltage-based_2013}.}
\end{algorithm}

\subsection{Power-based methods (MLP)}
The power-based method proposed in \cite{xu_phase_2018} is implemented. This uses Pearson correlation of the power measurements of each consumer and power measurements of each phase of the distribution transformer (\ref{eq:pearson_power}). To provide accurate results, some preprocessing of the power data is performed. Power variations between two timesteps $\Delta h_{i} = h_{i,t+1} - h_{i, t} \; \; \forall t$ are calculated. Secondly, salient components  $SH_{i}$ are extracted. These are the $m$ largest variations out of $T$ time samples. The corresponding salient variations of the transformer on each phase $j$ is denoted $SP_{j}$. 
In \cite{xu_phase_2018}, salient components that exceed a threshold are selected, as opposed to extracting the $m$ largest variations. However, the latter approach was found to be more robust with our data set and was therefore chosen in favor of the approach in \cite{xu_phase_2018}.
The implemented method is summarized in Algorithm 3.
\begin{equation}
    \rho(SH_{i},SP_{j}) = \frac{(SH_{i}-\mu(SH_{i})) (SP_{j}-\mu(SP_{j}))}{\sigma(SH_{i}) \sigma(SP_{j})} 
    \label{eq:pearson_power}
\end{equation}
In \cite{xu_phase_2018} a spectral analysis is also performed at the data preprocessing stage, which filters out the low frequency variations in the power data. Calculating \eqref{eq:pearson_power} on high frequency variations alone improves the results in \cite{xu_phase_2018}, but did not have a significant impact in our numerical illustrations. As such, spectral analysis is not discussed in this paper.
\begin{algorithm}[tb]
\SetAlgoLined
\DontPrintSemicolon
\KwData{$H_{i}=[h_{1,i},\ldots,h_{T,i}]$ for all consumers $i$\;
$P_{j}=[p_{1,j},\ldots,p_{T,j}]$  for each phase $j$ \;
} \;

 Sort consumer list $\ell$ according to highest $MAV_i = \frac{1}{T} \sum_{t=1}^{T} |h_{i,t}|$ of each consumer $i$ \;
 \For{Each phase $ j \in \{a,b,c\}$}{
    Calculate transformer power variations $\Delta P_{j} = [p_{1,j}-p_{2,j}, ... ,p_{T-1,j}-p_{T,j}]$ \;
    }
 \For{Each consumer $i \in \ell$}{
    Calculate power variations $\Delta H_{i} = [h_{1,i}-h_{2,i},\ldots, ,h_{T-1,i}-h_{T,i}]$ \;
    Select $m$ biggest variations $\Delta H_{i}$ and store as $SH_{i}$
    Select corresponding variations for $\Delta P_{j}$ for each $j \in \{a,b,c\}$ and store as $SP_{i,j}$ \;
    \For{Each phase $ j \in \{a,b,c\}$}{
        $c_{i,j} \leftarrow \rho(SH_{i},SP_{j})$ using (\ref{eq:pearson_power}) \;
        }
Allocate consumer $i$ to phase $j$ with highest  $c_{i,j}$ \;
Subtract power profile from transformer measurement $P_{j} \leftarrow P_{j} - H_{i}$
}
\caption{MLP \cite{xu_phase_2018}}
\end{algorithm}

In the MLP algorithm, the consumers with the highest variability in their time-series measurement, according to the mean average variability (MAV) are assigned to one of the transformer phases first. This is because consumers with higher variability are more likely to be allocated to the correct phase, since they contain more distinct changes in power, which might correspond to turning on a big electrical appliance. Therefore, they tend to show higher correlation with the transformer's salient components.
\subsection{Ensemble methods}
Two novel ensemble learning methods are proposed based on two common types of ensembles: `bagging' and `boosting'.
\subsubsection{Bagging ensemble method}
\begin{figure}[t]
    \centering
    \includegraphics[width=\linewidth]{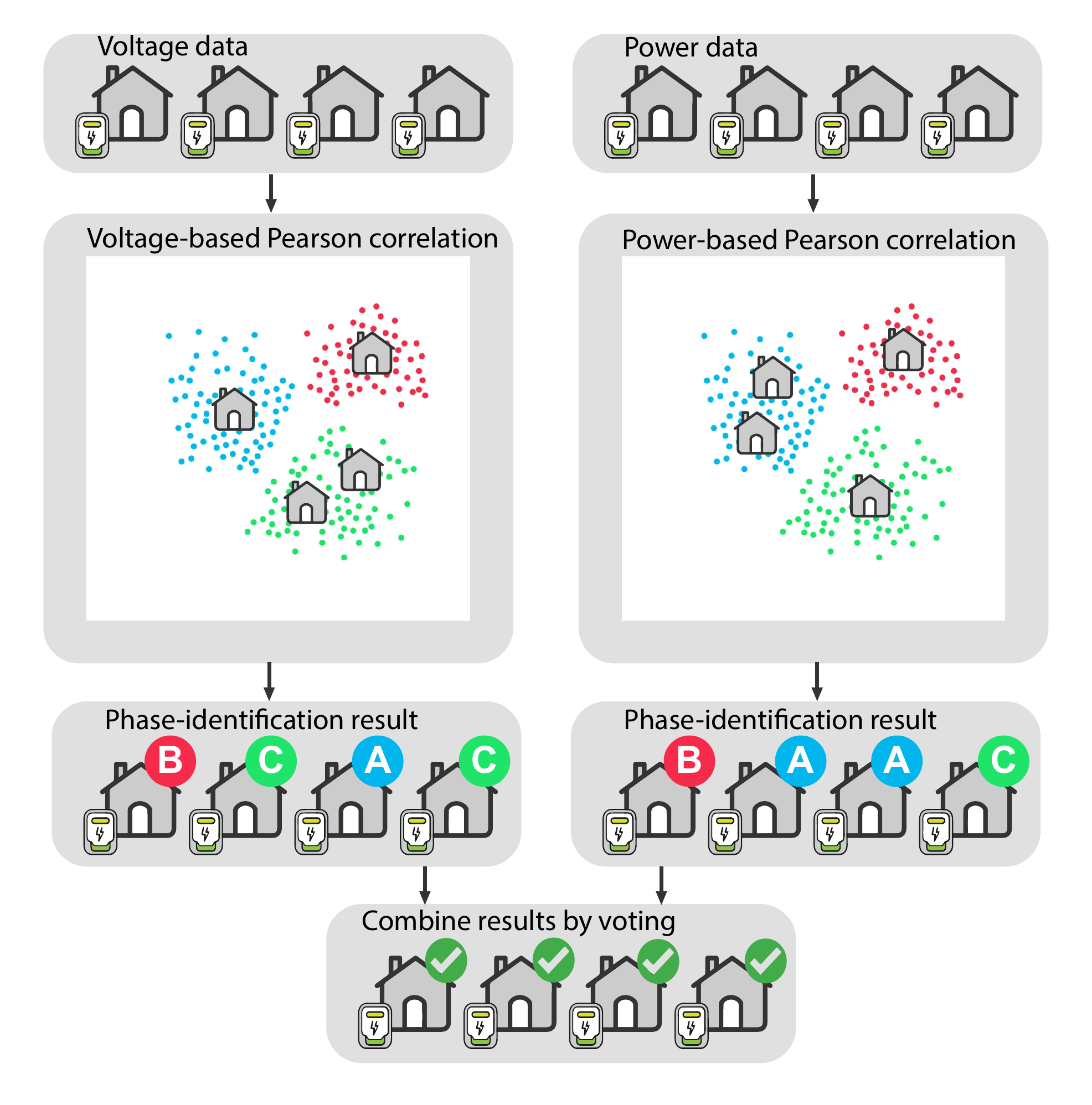}
    \caption{Bagging ensemble with voltage and power data.}
    \label{fig:bagging}
\end{figure}

 In this method, different data sources are used with the same algorithm, and the respective results are combined according to a chosen combination metric. In a variety of other machine learning applications, such approach leads to improved results~\cite{sagi_ensemble_2018}. Fig.~\ref{fig:bagging} illustrates how such an approach could be applied to phase identification. Voltage and power data can be separately collected and used by different methods to obtain a phase identification result. The results are then combined, e.g. by majority voting.
\one{A simple variant of bagging ensemble is implemented in this work, as a reference to compare the proposed boosting ensemble against.} The implemented bagging reference method uses the result from the voltage-based technique and if voltage data is not available of a certain customer, it completes the phase identification with the power-based technique. This approach, although simple, leads to good results in the analyzed numerical illustrations.

\subsubsection{Boosting ensemble method}
\begin{figure}[t]
    \centering
    \includegraphics[width=\linewidth]{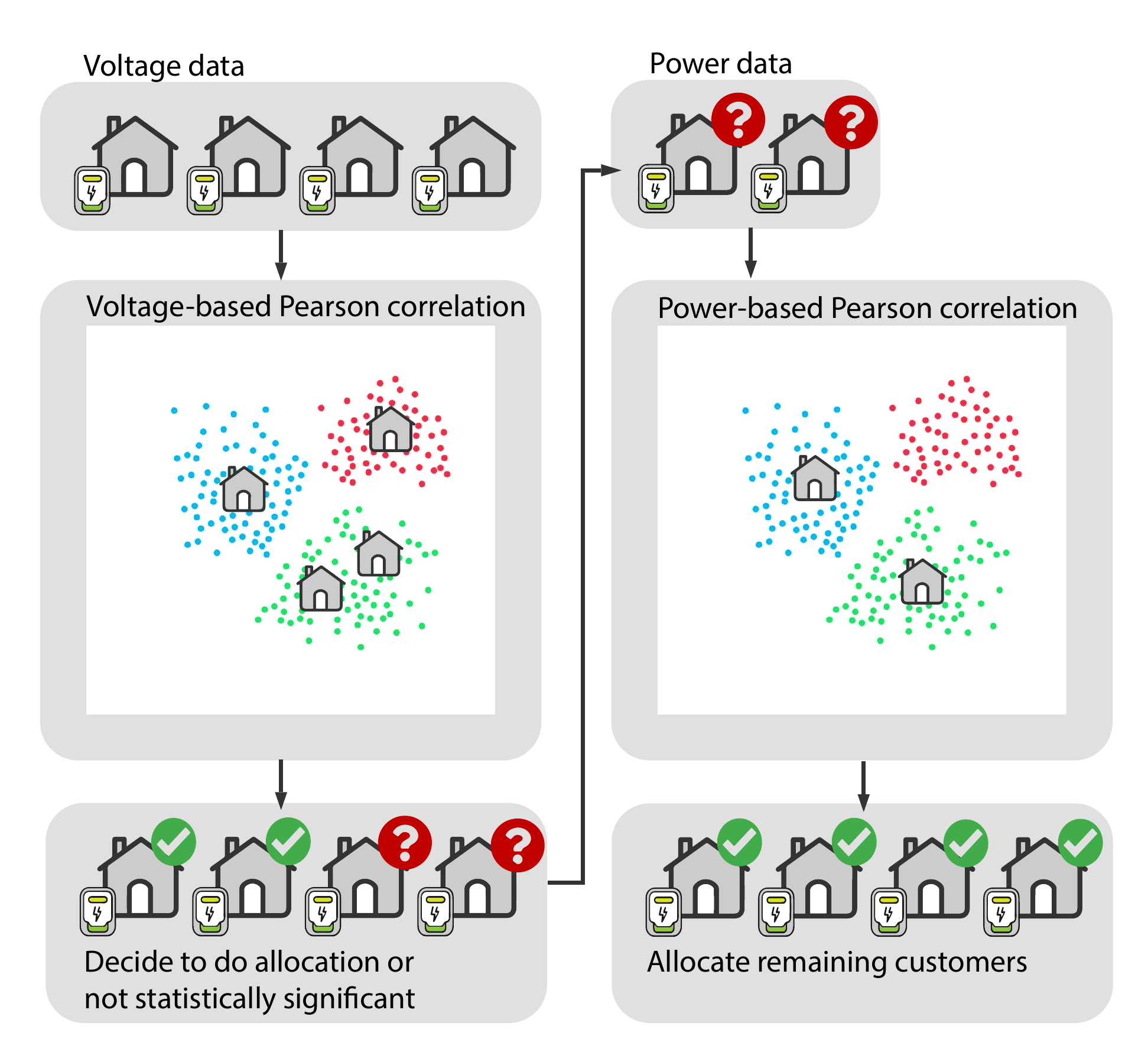}
    \caption{Boosting ensemble with voltage and power data.}
    \label{fig:boosting}
\end{figure}

In this method, a second model is trained using the data that are wrongly classified by the first model. However, this requires labelled data and therefore does not lend itself well to phase identification, as phase labels that are not identified by the first method remain unavailable. Therefore, a variant is implemented that does the phase identification partially using voltage correlation. This technique is only applied to those consumers for which this decision can be made with sufficient confidence. \three{ A minimum threshold $\rho_{threshold}$ is used on the voltage correlation. In our work this threshold is set at $0.2*T$, with $T$ being the length of the time-series. This was determined by inspecting the distribution of the obtained voltage Pearson correlation coefficients. By setting the threshold at $0.2*T$, there is a clear distinction between the highest correlating and second to highest correlation phase, for most feeders. A more rigorous method to determine the threshold or an alternative way to determine which method to use could be the scope of future work. The threshold is dependent on the input length $T$ to scale the threshold appropriately in the numerical illustration where input length is varied.} The full method is illustrated in Fig.~\ref{fig:boosting}.
Compared to the bagging ensemble method, the power-based step of the boosting ensemble method is expected to perform better. This is because the demand of the users which have their phase connectivity identified with sufficient confidence by the voltage method, are subtracted from the transformer power measurement. The MLP algorithm uses the residue to identify the remaining users.
\begin{algorithm}[htb]
\SetAlgoLined
\DontPrintSemicolon
\KwData{$U_{i}=[u_{1,i},\ldots,u_{T,i}]$ for all consumers $i$, \;
\indent $H_{i}=[h_{1,i},\ldots,h_{T,i}]$ for all consumers $i$\;
\indent $U_{ref,j}=[u_{ref,1,j},\ldots,u_{ref,T,j}]$ for each phase $j$ \;
\indent $P_{j}=[p_{1,j},\ldots,p_{T,j}]$  for each phase $j$} \;
 \For{Each consumer $i \in \{1,\ldots,N \}$}{
    \For{Each phase $ j \in \{a,b,c\}$}{
    Calculate $\rho(U_{i},U_{ref,j})$ using (\ref{eq:pearson}) \;
    }
    \eIf{max($\rho(U_{i},U_{ref,j}), \forall $j$) > \rho_{threshold}$}{
        Allocate $i$ with highest  $\rho(U_{i},U_{ref,j})$ to phase $j$\;}
    {
        Use MLP method (Algorithm 3) to allocate customer $i$\;}
}
\caption{Boosting ensemble}
\end{algorithm}

\section{Numerical illustration}
\label{sec:numill}
Three numerical illustrations have been analyzed to benchmark the different methods. Realistic assumptions are made on the measurements, so that they resemble those that are typically available to a DSO in a practical case. 
\subsection{Test feeders}
\begin{figure}[b]
    \centering
    \includegraphics[width=\linewidth]{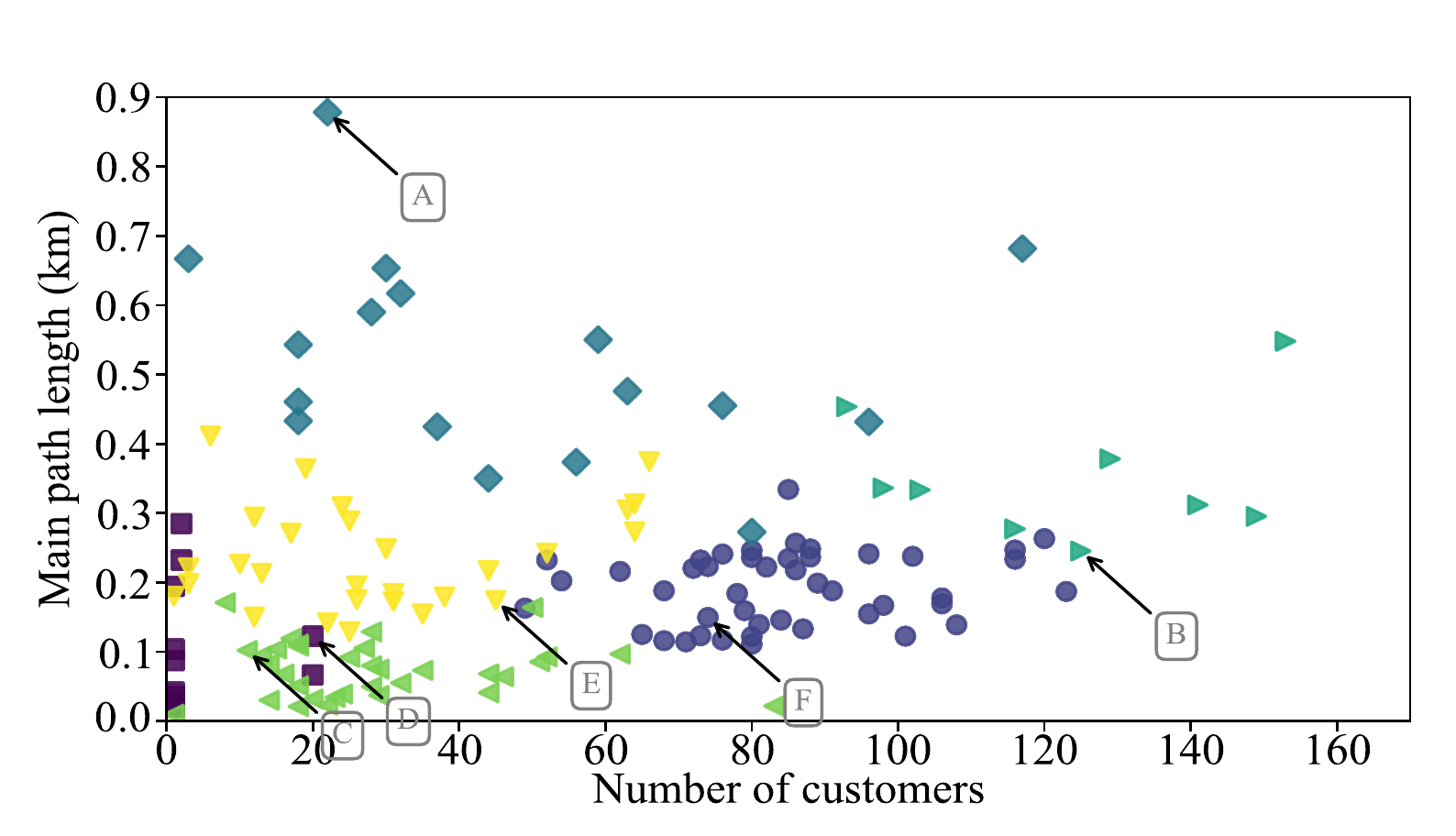}
    \caption{Clusters and selected representative feeders from the non-synthetic European DS \cite{koirala_non-synthetic_2020}.}
    \label{fig:repr_feeders}
\end{figure}

A non-synthetic data set of a European DS is used \cite{koirala_non-synthetic_2020}, which contains real SM power measurements and network topology information. From the 160 radial LV feeders in this data set, six representative feeders are selected according to the approach in \cite{rigoni_representative_2016}, which clusters the feeders into groups with similar features and then extracts one feeder from each cluster. The selected feeders' properties are visualized in Figure~\ref{fig:repr_feeders} and summarized in Table \ref{tab:repr_feeders}. \three{In this work, a radial LV distribution network is studied. In principle, nothing prevents the application of the  methods to meshed LV grids. In practice, voltage-based results might be worse, as voltage variations might be less evident on meshed networks. 
However, the meshed LV networks data is scarce, and we do not perform this analysis for such network in the paper.}
\begin{table}[tb]
    \centering
     \caption{Features of the selected representative feeders.}
    \begin{tabular}{ccccc}
    \hline
    \textbf{Case}&\textbf{Nb of} & \textbf{Yearly energy} & \textbf{Main path}  & \textbf{Average path}  \\
    &\textbf{users}& \textbf{per user (kWh)} & \textbf{length (m)} & \textbf{impedance ($\Omega$)} \\
    \hline
   A& 22& 1852 & 878&
 0.206\\
 B & 125& 1894 & 245&
 0.117\\
 C & 11& 1802 & 102&
 0.0598\\
 D & 20& 4173 & 138&
 0.196\\
E & 45 & 2066 & 173&
       0.106\\
 F & 74& 1905 & 149&
 0.0914\\
    \hline
    \end{tabular}
   
    \label{tab:repr_feeders}
\end{table}

\subsection{Benchmarking existing phase identification methods}
\begin{figure}[t] 
  \begin{subfigure}[b]{0.48\linewidth}
    \centering
    \includegraphics[width=\linewidth]{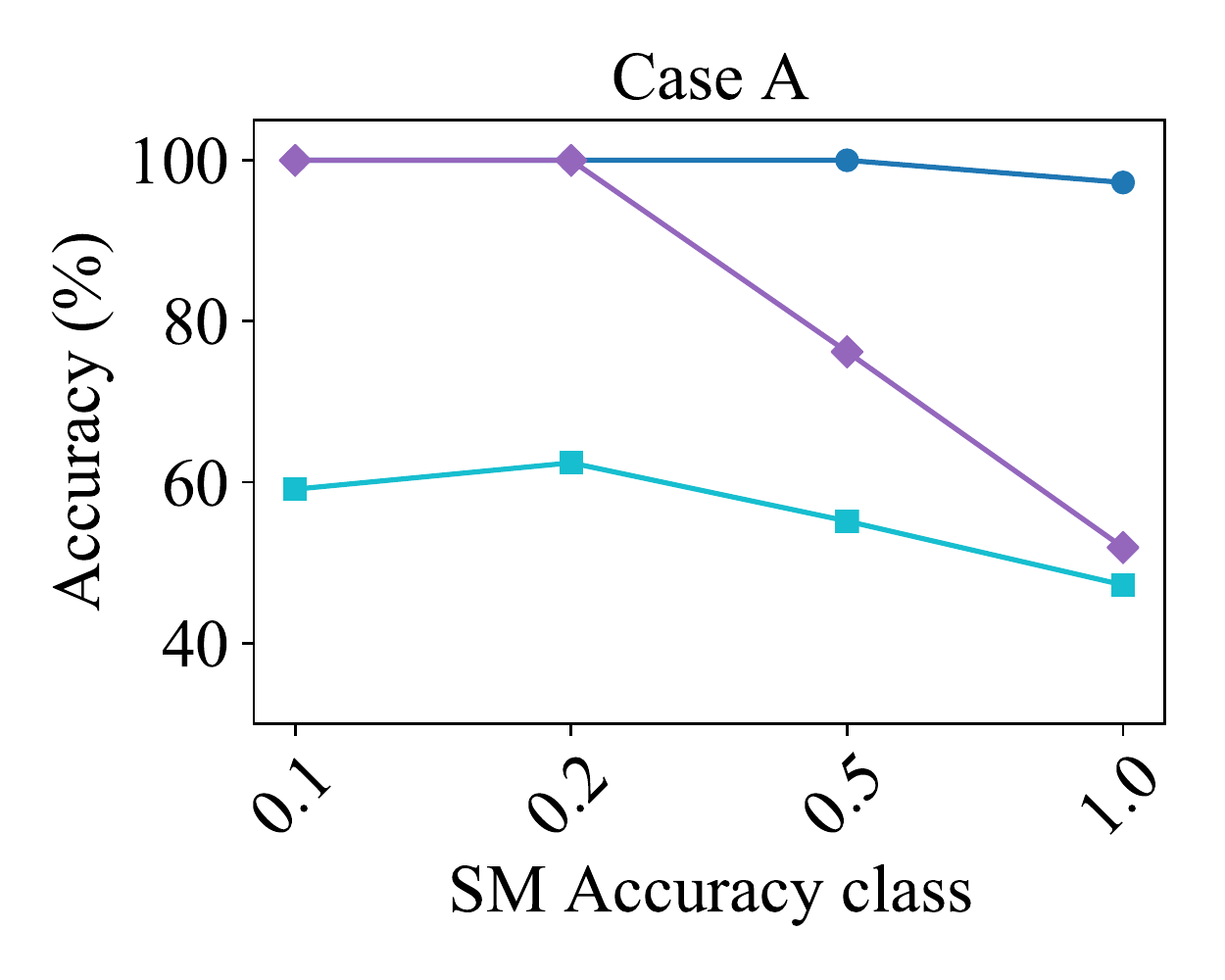}
    \label{fig7:a} 
  \end{subfigure}%
  \vspace{-5mm}
  \begin{subfigure}[b]{0.48\linewidth}
    \centering
    \includegraphics[width=\linewidth]{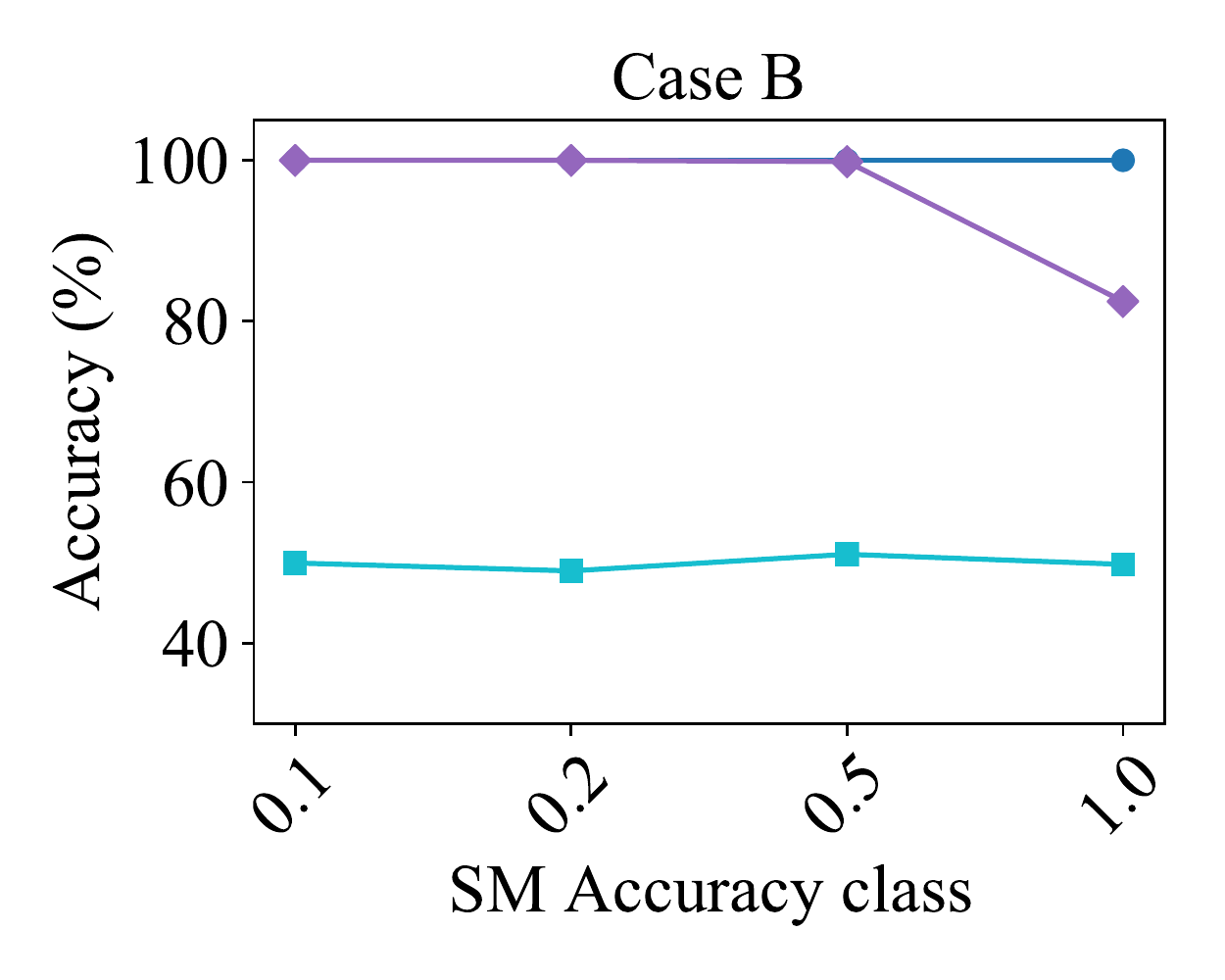}
    \label{fig7:b} 
  \end{subfigure}
  \begin{subfigure}[b]{0.48\linewidth}
    \centering
    \includegraphics[width=\linewidth]{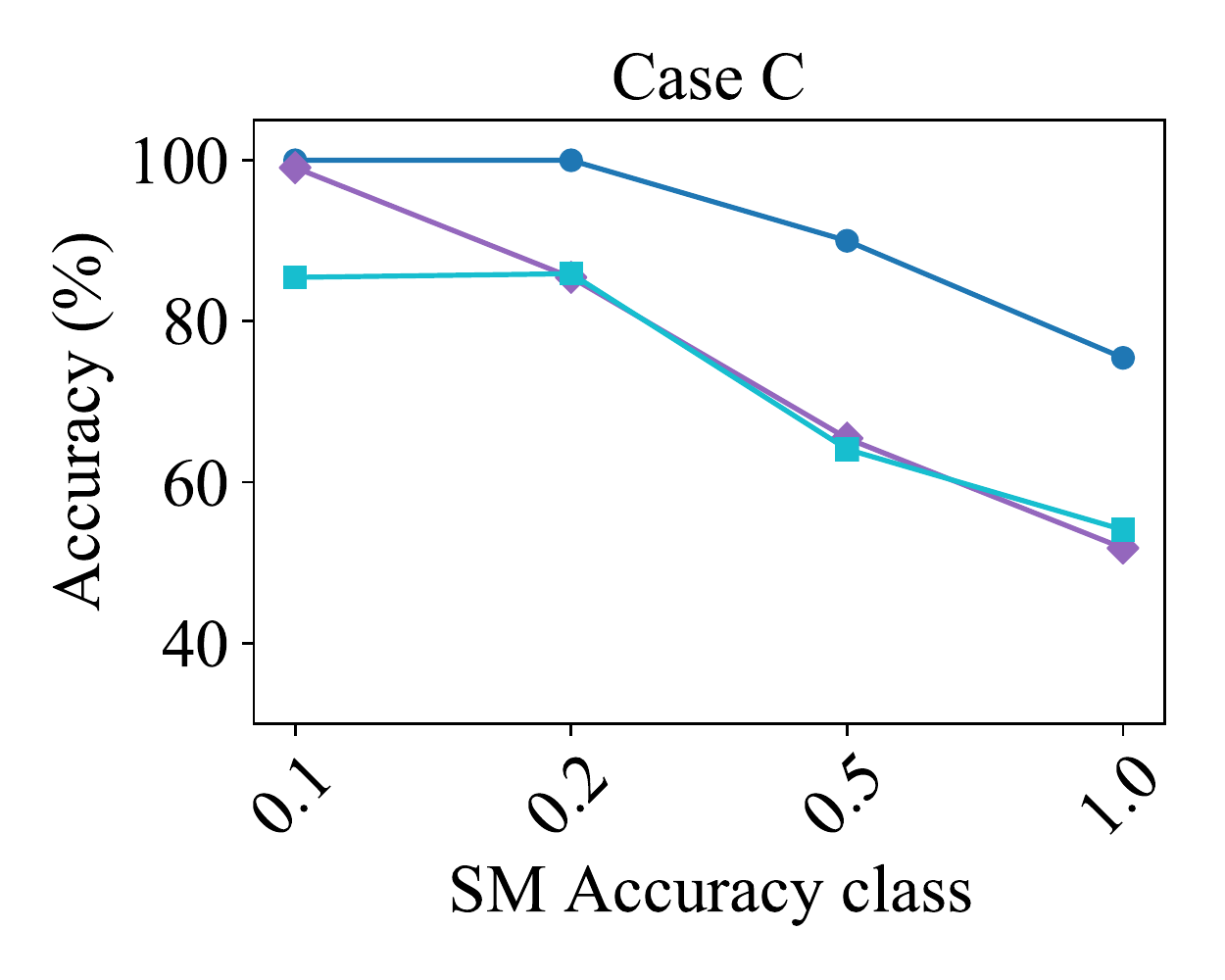}
    \label{fig7:c} 
  \end{subfigure}%
    \vspace{-5mm}
  \begin{subfigure}[b]{0.48\linewidth}
    \centering
    \includegraphics[width=\linewidth]{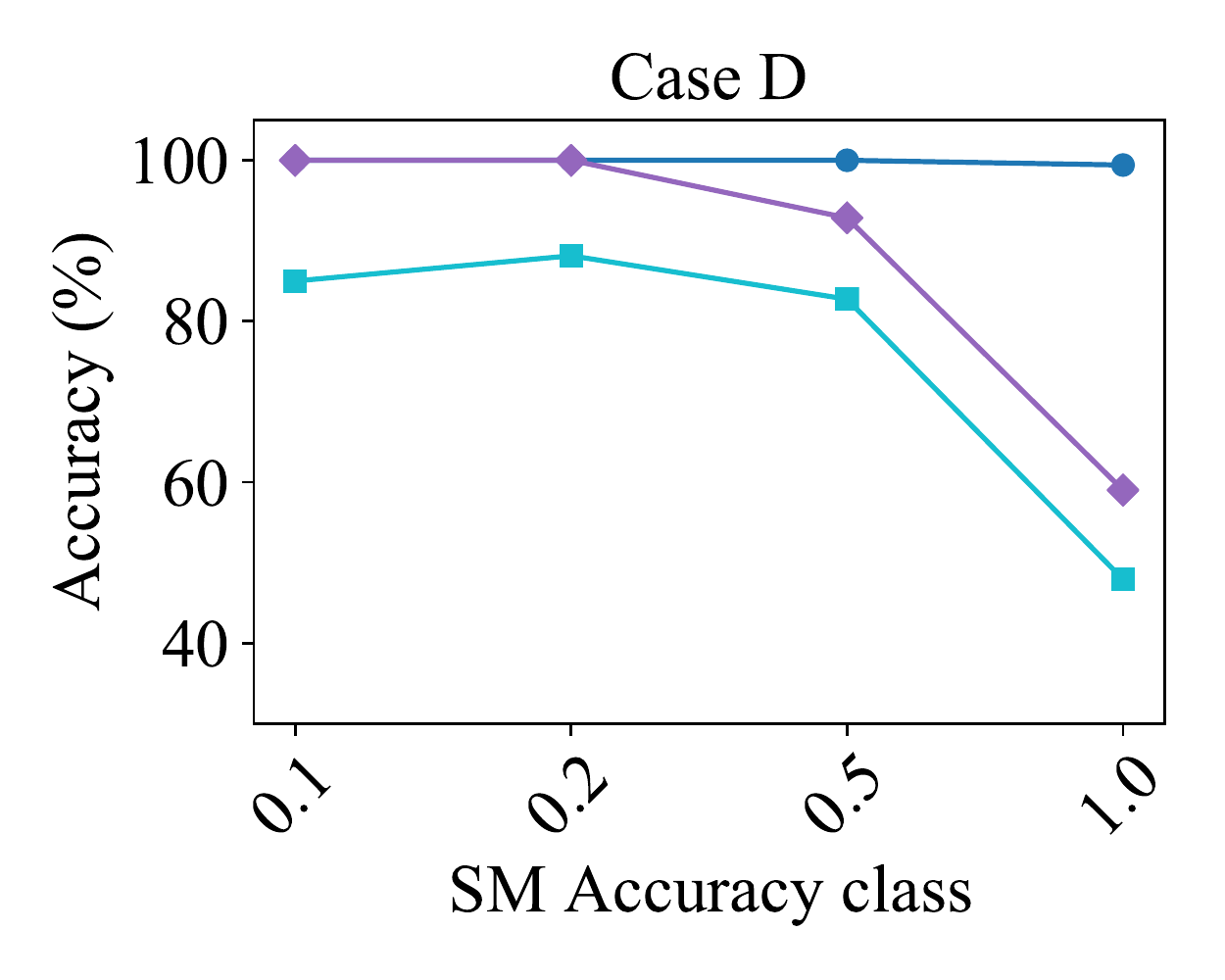}
    \label{fig7:d} 
  \end{subfigure}
    \begin{subfigure}[b]{0.48\linewidth}
    \centering
    \includegraphics[width=\linewidth]{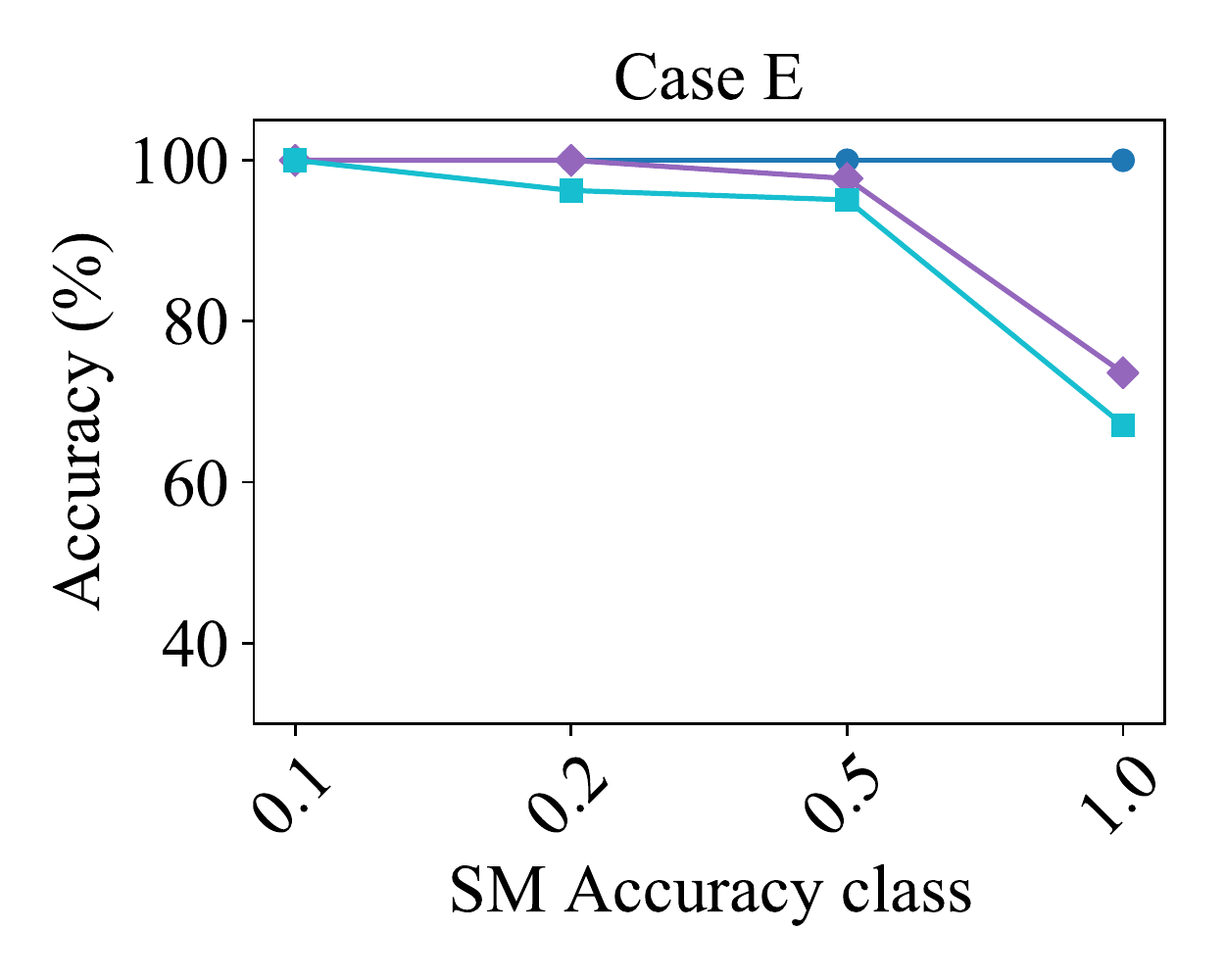}
    \label{fig7:c} 
  \end{subfigure}%
  \begin{subfigure}[b]{0.48\linewidth}
    \centering
    \includegraphics[width=\linewidth]{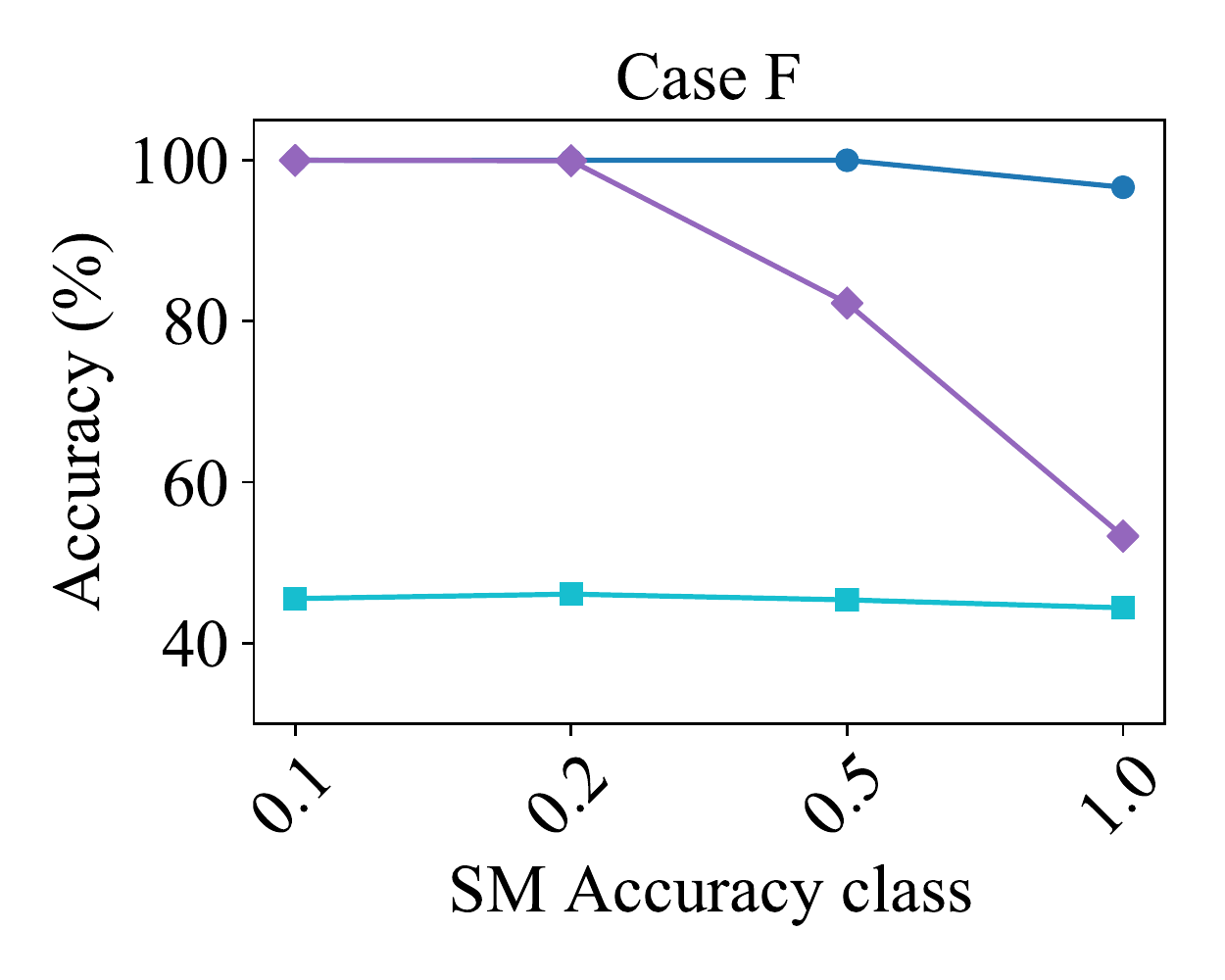}
    \label{fig7:d}
  \end{subfigure}
  \centering
    \includegraphics[width=0.7\linewidth]{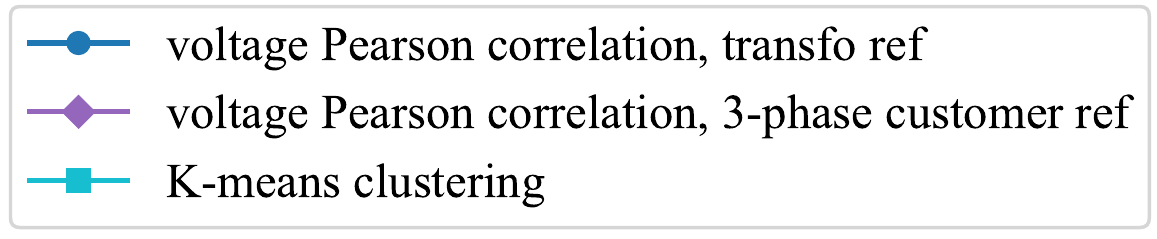}
  \caption{Accuracy for voltage based phase identification methods for all SM classes in the six representative feeders.}
  \label{fig:method_comparison_volt} 
\end{figure}
In this numerical illustration, phase identification is performed with a complete dataset, i.e., voltage and power measurements time-series of equal time length, time resolution and accuracy are used. First, the deterministic Pearson correlation method \cite{pezeshki_correlation_2012} and K-means clustering method \cite{arya_voltage-based_2013} from the literature are compared to each other and to a variant of the voltage-based Pearson correlation which uses a three-phase customer as the reference instead of the transformer \cite{pezeshki_consumer_2012}. The results for each representative feeder and accuracy class are shown in Fig. \ref{fig:method_comparison_volt}. The accuracy on the y-axis shows the fraction of the customers that were allocated to the correct phase by the algorithm.

Fig. \ref{fig:method_comparison_volt} shows that for the correlation approaches it is better to use the transformer as a reference, rather than a three-phase customer. A possible explanation for this is that customers that are on a feeder branch which is different from that of the reference user, correlate worse to the reference user's voltage pattern. This could happen as voltage variations in a branch are mainly influenced by the  current through the branch sections, whereas the transformer voltage variations account for the total feeder currents. As such, the transformer voltage appears to be a good reference for all the feeders' users.

In addition to the non-synthetic DS data \cite{koirala_non-synthetic_2020}, the methods are also applied to a second dataset \cite{commission_for_energy_regulation_cer_notitle_2012}. When applying the phase identification algorithms on this dataset, a higher accuracy is obtained, similar that reported in \cite{xu_phase_2018}. The results are omitted here but this shows however, that the results of MLP methods depend on the used data set, which is generally not the case for MLV methods.
 
\begin{table}[b]
 \caption{Obtained accuracy for all SM classes and methods, in average for the 6 cases. Best results for every SM class are highlighted. Bagging ensemble is omitted since results are identical to volt. Pearson.}
    \centering
    \begin{tabular}{p{0.18\textwidth}cccccc}
    \hline
    \textbf{Method}&\textbf{0.2s}&\textbf{0.5s} & \textbf{0.1} & \textbf{0.2}  & \textbf{0.5} & \textbf{1.0}  \\
    & \textbf{(\%)}& \textbf{(\%)}& \textbf{(\%)}& \textbf{(\%)}& \textbf{(\%)}& \textbf{(\%)}\\
    \hline
   volt. Pearson, transfo ref & \textbf{100} & 97.2 & \textbf{100} & \textbf{100} & \textbf{97.2} & \textbf{95.0} \\
 volt. Pearson, customer ref & \textbf{100} & 86.3 & \textbf{100} & \textbf{100} & 86.3 & 62.8 \\
 K-means clustering & 71.0 & 65.3 & 70.2 & 71.0 & 65.3 & 52.5 \\

 power Pearson & 97.2 & 93.3 & 79.8 & 62.3 & 44.0 & 39.1 \\


boosting ensemble & \textbf{100} & \textbf{100} & \textbf{100} & 98.6 & 93.7 & 83.5 \\

    \hline
    \end{tabular}
  
    \label{tab:results}
\end{table}
\begin{figure}[t] 
  \begin{subfigure}[b]{0.49\linewidth}
    \centering
    \includegraphics[width=\linewidth]{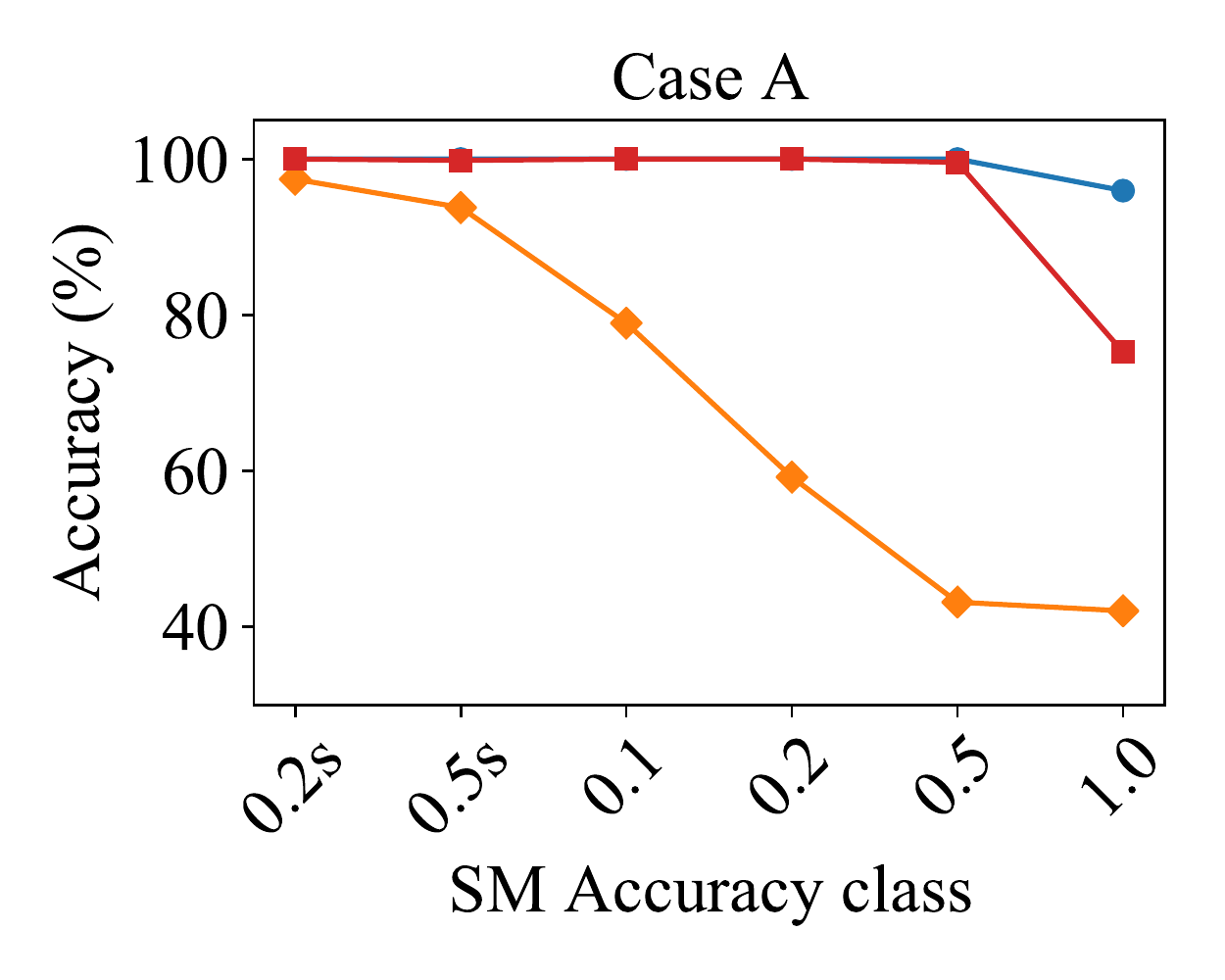}
    \label{fig7:a} 
  \end{subfigure}%
  \vspace{-5mm}
  \begin{subfigure}[b]{0.49\linewidth}
    \centering
    \includegraphics[width=\linewidth]{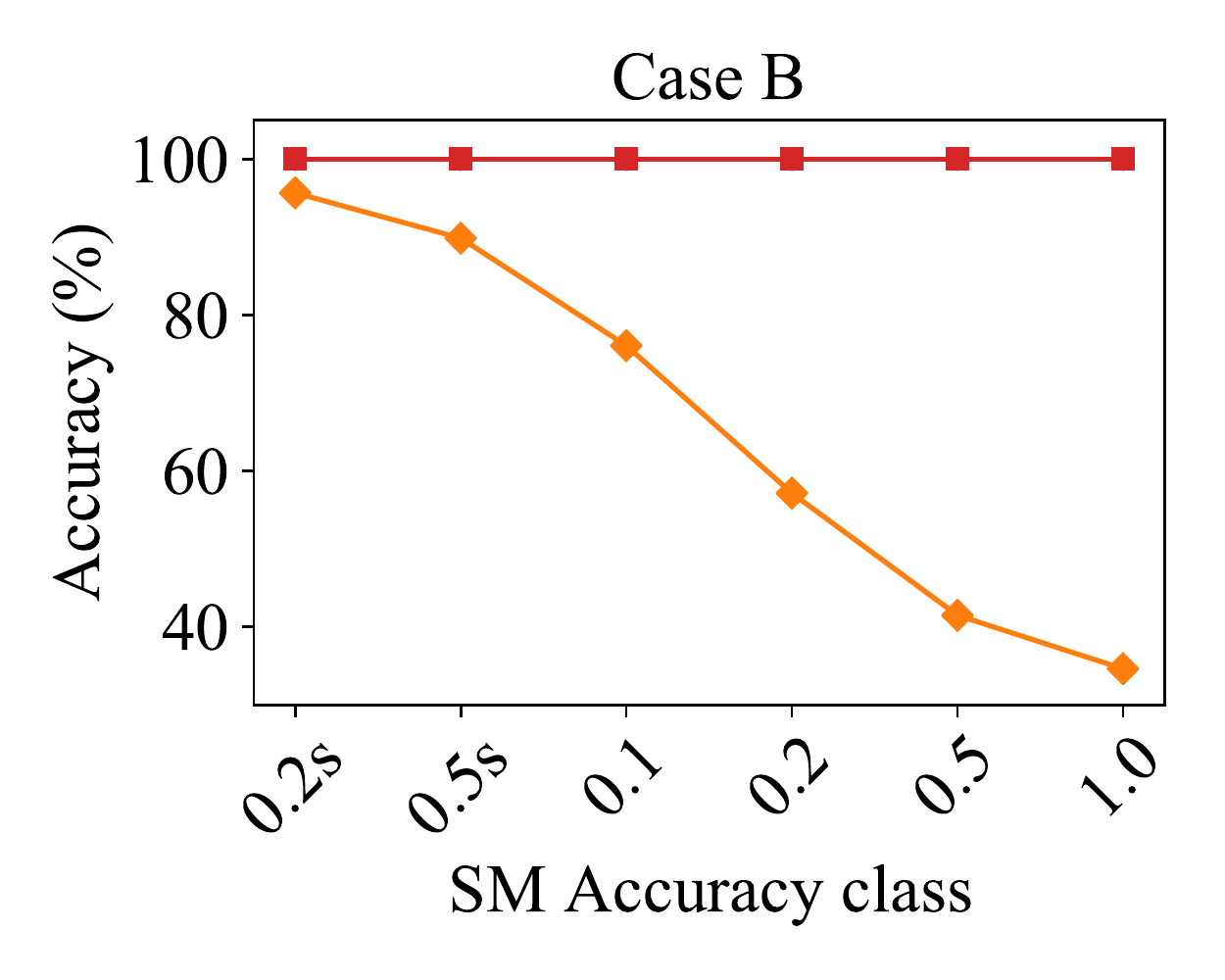}
    \label{fig7:b} 
  \end{subfigure}
  \begin{subfigure}[b]{0.49\linewidth}
    \centering
    \includegraphics[width=\linewidth]{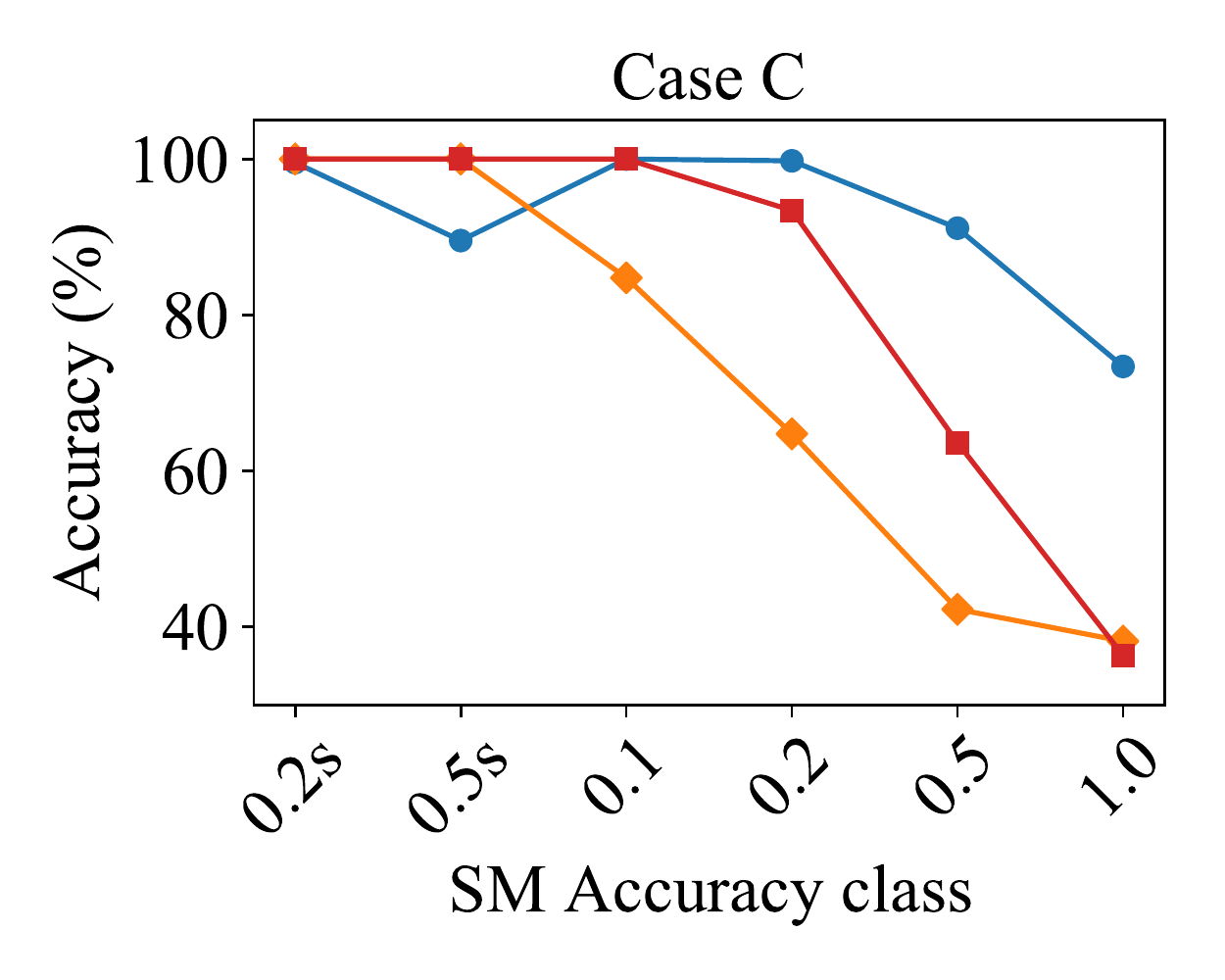}
    \label{fig7:c} 
  \end{subfigure}%
    \vspace{-5mm}
  \begin{subfigure}[b]{0.49\linewidth}
    \centering
    \includegraphics[width=\linewidth]{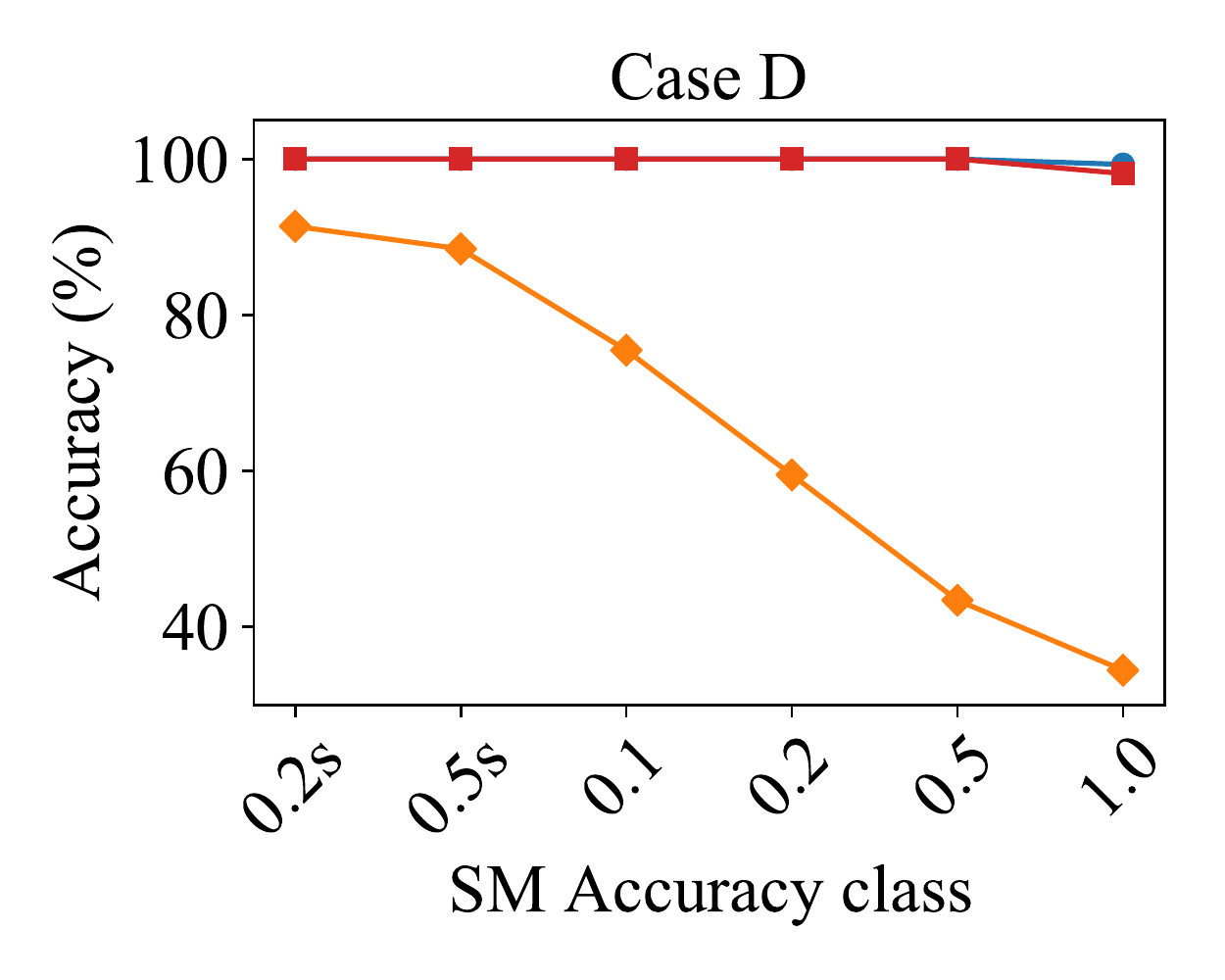}
    \label{fig7:d} 
  \end{subfigure}
    \begin{subfigure}[b]{0.49\linewidth}
    \centering
    \includegraphics[width=\linewidth]{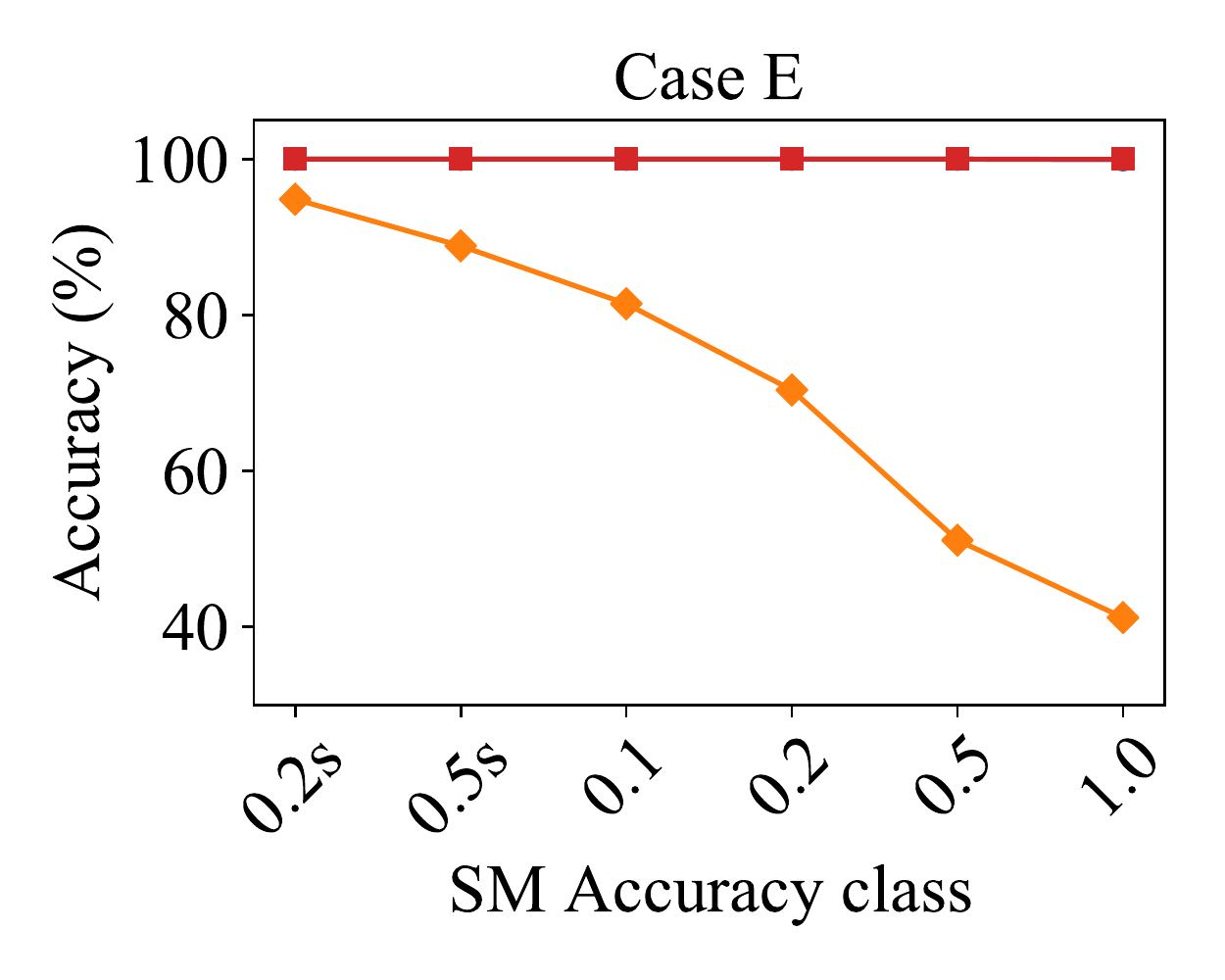}
    \label{fig7:c} 
  \end{subfigure}%
  \begin{subfigure}[b]{0.49\linewidth}
    \centering
    \includegraphics[width=\linewidth]{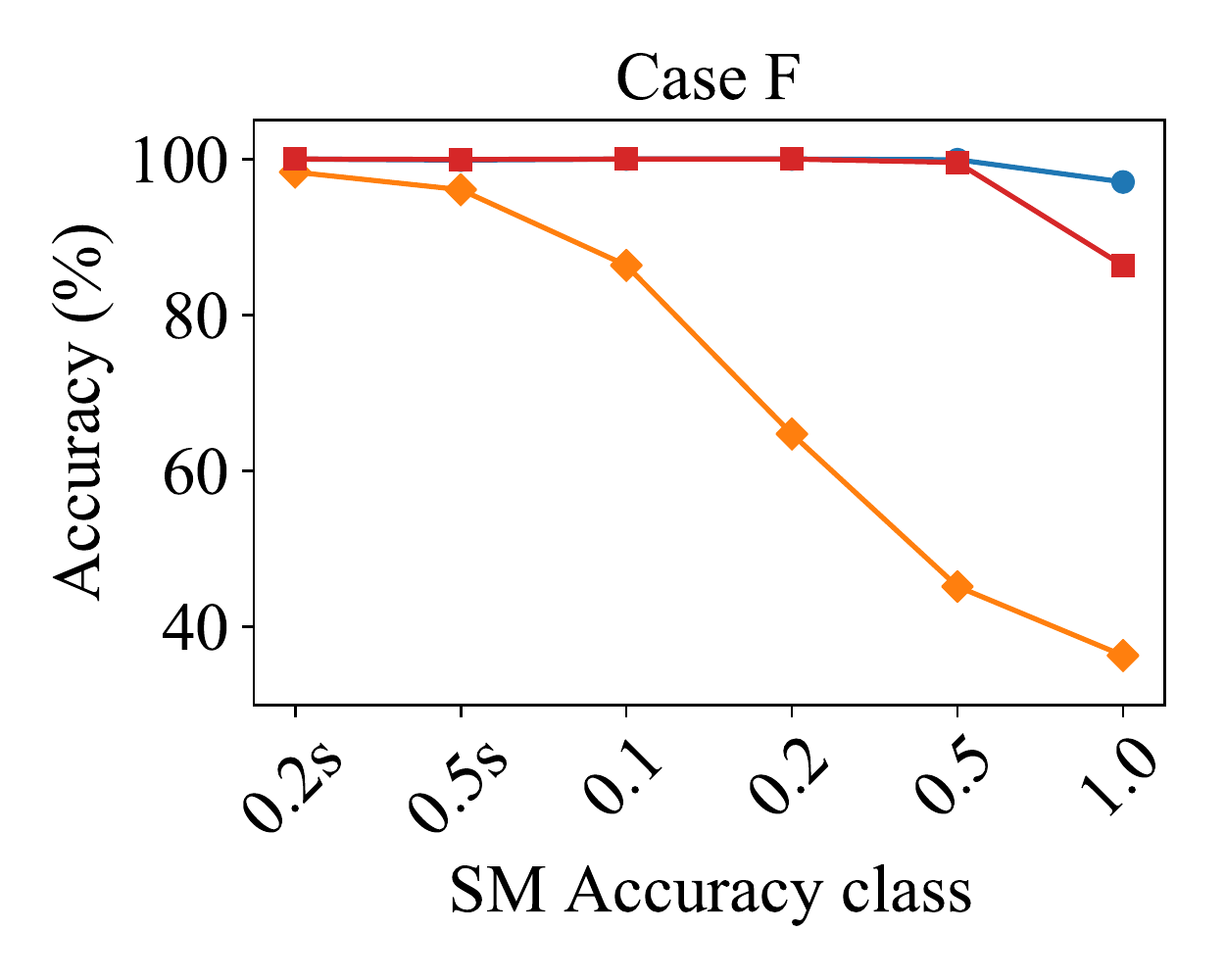}
    \label{fig7:d}
  \end{subfigure}
  \centering
    \includegraphics[width=.5\linewidth]{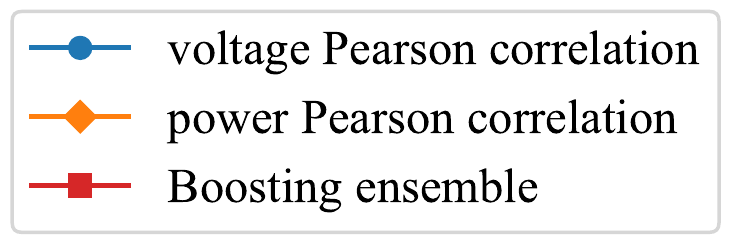}
  \caption{Accuracy of ensemble and power-based methods compared to voltage-based methods for all SM classes in the six representative feeders. Bagging ensemble results are identical to volt. Pearson.}
  \label{fig:method_comparison} 
\end{figure}

In Figure \ref{fig:method_comparison} \two{and Table \ref{tab:results} , the ensemble methods are compared with the voltage-based method using the transformer reference and with power-based algorithms.} The comparison is made based on equal quality and quantity of data. Voltage-based techniques outperform power-based techniques in almost all cases and accuracy classes. The power-based technique outperforms the voltage-based method in only one case, i.e., 0.5s class SM in case C. In this case, the boosting ensemble method can identify that it cannot allocate consumers using the voltage-based technique with high confidence. Instead, it correctly allocates the remaining consumers with the power-based technique and obtains an accuracy of 100\%. The boosting ensemble uses the results of the power-based technique even if (relatively) low accuracy meters such as 1.0 class are used. Such low accuracy meters usually yield poor results when using a power-based technique. This can be seen in case A and case C. The bagging ensemble is omitted since it would yield the same results as the voltage-based method in the case of complete data.

\subsection{Impact of reduced voltage measurement points}
As a second case, a numerical illustration is studied where voltage data for 20 days is only available for a limited number of consumers, while power data is available for all consumers for 20 days. \three{This could be the case in real-life if power time-series are always available for billing purposes while supplementary voltage data are collected for phase identification or other grid-management applications. It could happen that only part of the consumers allow for their voltage data to be shared with the DSO because of data protection rules, or because they have newer generation SM.}

\begin{figure}[tbh]
    \centering
    \begin{subfigure}[b]{\linewidth}
    \centering
    \includegraphics[width=\linewidth]{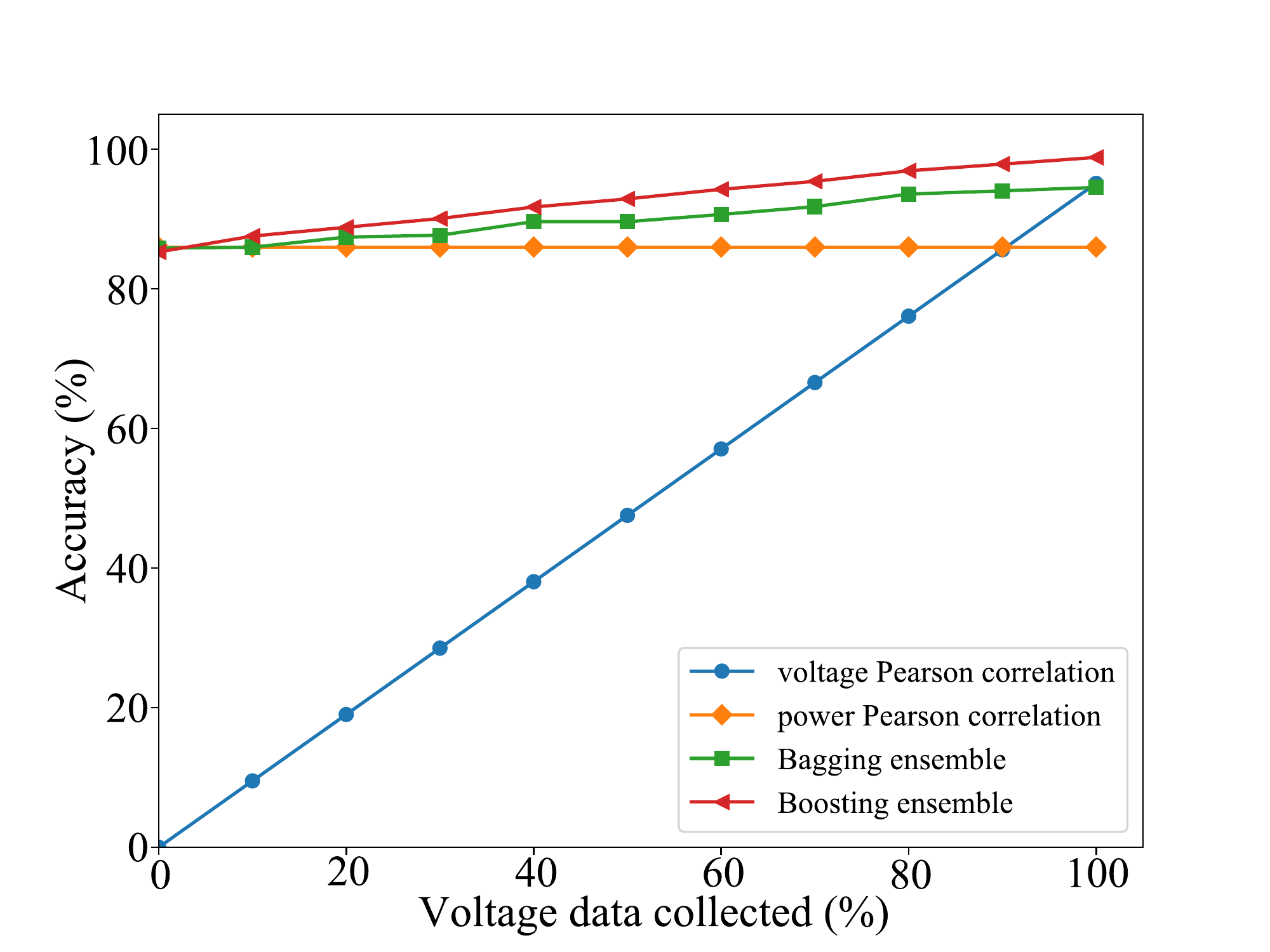} 
    \caption{Class 0.5s SM}
        \label{fig:8a} 
  \end{subfigure}
  \begin{subfigure}[b]{\linewidth}
    \centering
    \includegraphics[width=\linewidth]{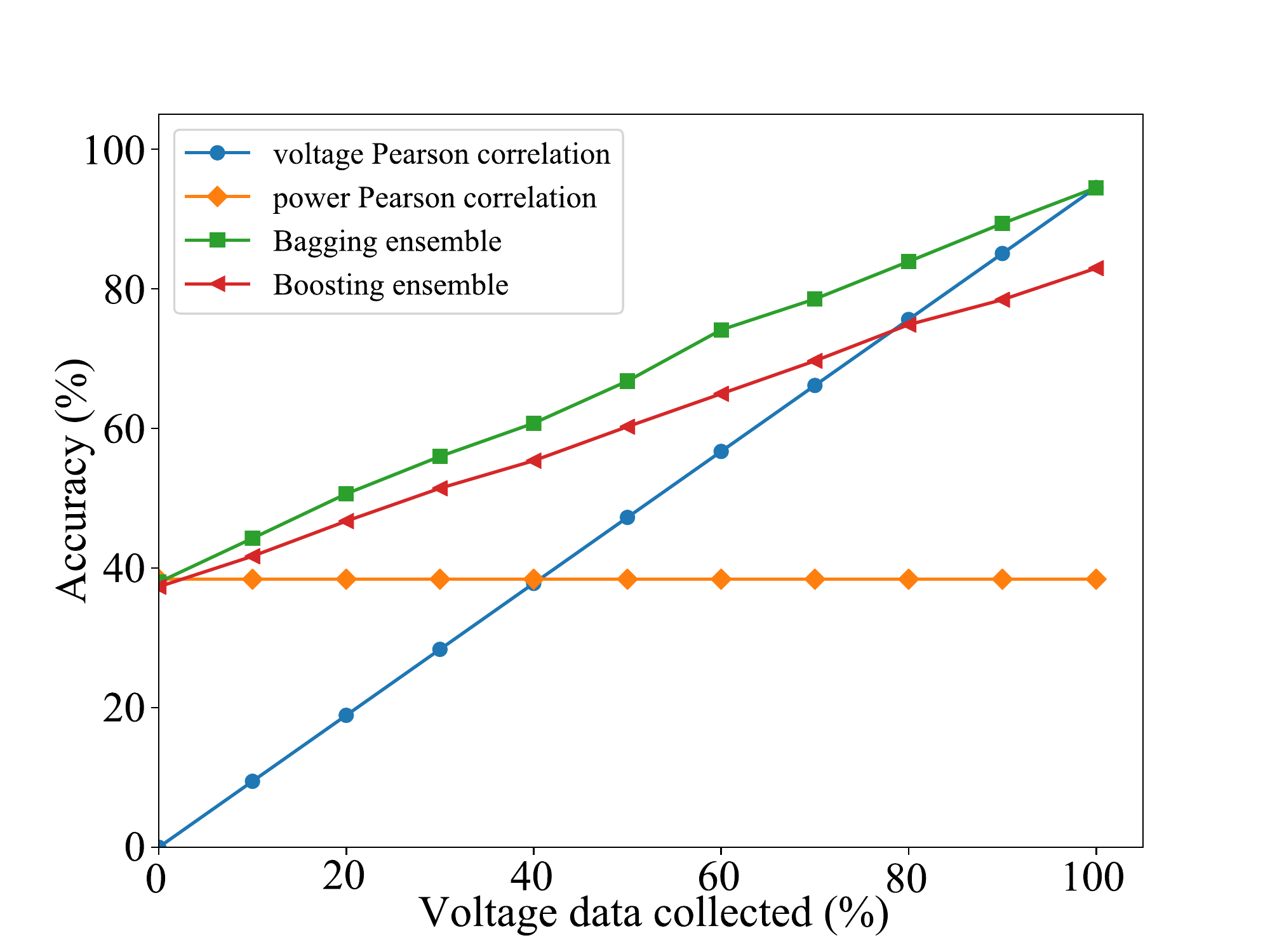}
    \caption{Class 1.0 SM}
        \label{fig:8b} 
  \end{subfigure}
  \caption{Influence of fraction of consumers for which 20 day voltage data is collected on accuracy.}
  \label{fig:8}
  \end{figure}

In Fig.~\ref{fig:8}, the accuracy is shown in function of the fraction of consumers for which 20 days of voltage data are collected. The SM accuracy class displayed are 0.5s and 1.0. The accuracy obtained by the voltage-based method improves linearly with the fraction of consumers with available data, and reaches 97.2\% in the case of 0.5s class SM and 95.0\% in the case of a 1.0 class SM. The power-based method is invariant to the change in voltage data collected and therefore obtains higher accuracy than the voltage-based approach at low levels of voltage data collected but worse at high levels. The boosting ensemble can outperform the simple bagging approach when 0.5s class SM is used, as shown in Fig.~\ref{fig:8a}. However, when a 1.0 class SM is used, the boosting method  obtains worse results than the bagging ensemble, as can be seen in Fig.~\ref{fig:8b}. One possible reason for this is that the tuning of the threshold was done based on the Pearson correlation distribution obtained using a 0.5s class SM.
\subsection{Impact of reduced voltage measurement length}
In this numerical illustration, a situation is studied where power data is available for 20 days, and supplementary voltage data is collected for a limited duration from all customers. This could be the case in practice if the network has a high SM penetration level, but the DSO wants to limit the amount of supplementary voltage data to collect and store.

\begin{figure}[htb]
    \centering
    \includegraphics[width=\linewidth]{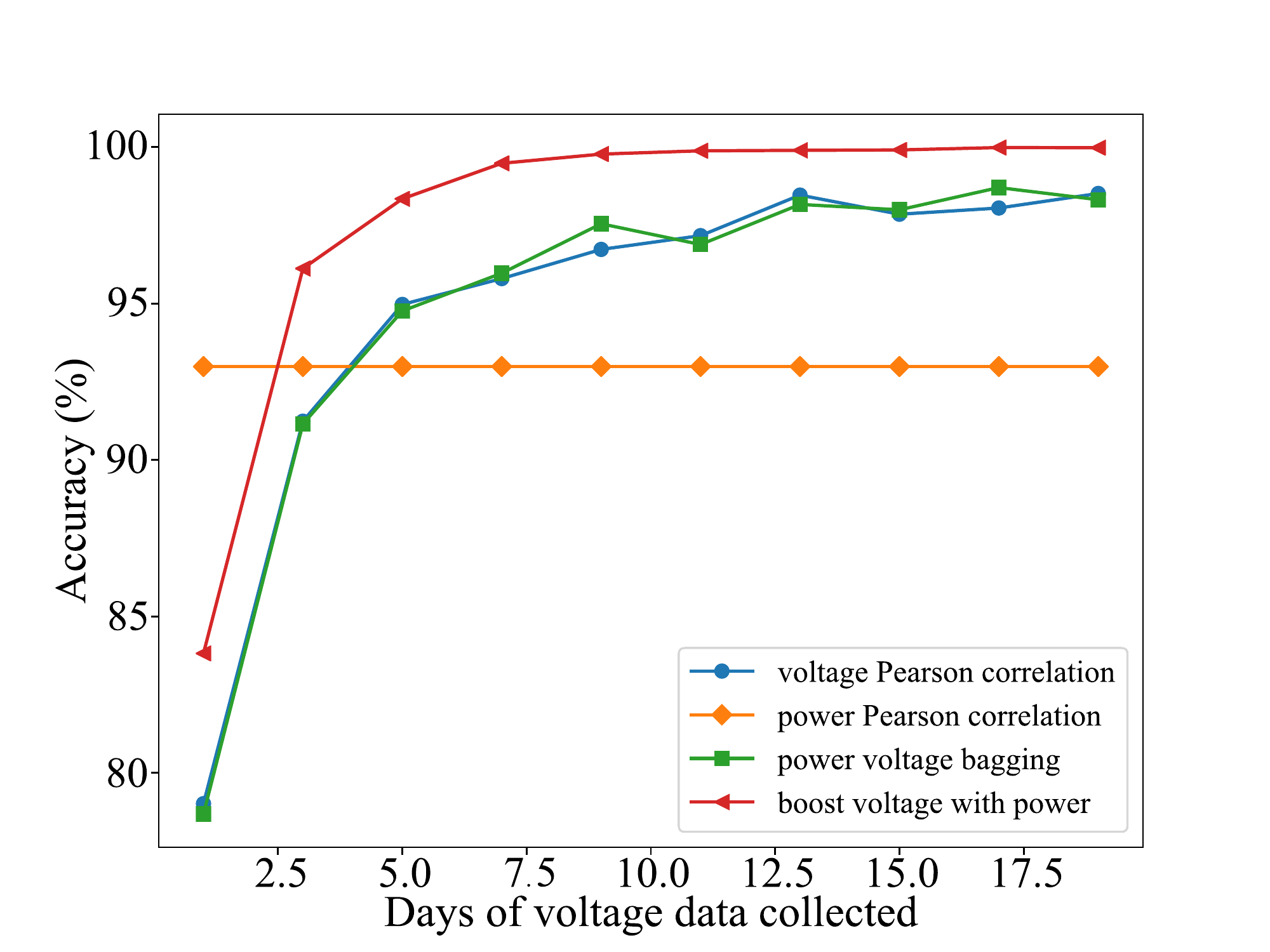}
    \caption{Accuracy for varying quantity of voltage data collected and 20 days of power data  (Class 0.5s SM).}
    \label{fig:spain}
\end{figure}

In Fig.~\ref{fig:spain}, the obtained results are shown using 20 days of power data and between 2 and 20 days of voltage data from a 0.5s class SM. This analysis shows that the boosting ensemble can improve the average accuracy of the studied feeders by 2.8\%, compared to the voltage-based method. It can be observed that with limited voltage data, the boosting ensemble still overestimates the accuracy of the voltage method and, as a result, is outperformed by the MLP. It can also be seen that the accuracy saturates with increasing days of collected voltage data, especially in the case of the boosting ensemble.

The boosting approach only outperforms the voltage-based method when a 0.5s class SM is used. In the case of a higher 0.2s class SM, the voltage-based method already obtains 100\% and in all lower class SM, the results of the power-based method are low to start with, which makes it hard for the ensemble methods to improve upon the voltage-based result.
\subsection{Result Analysis}
When benchmarking the existing methods, the voltage-based Pearson correlation method showed better results compared to the use of K-means \one{(see Figure \ref{fig:method_comparison_volt})}. It allocated all consumers correctly in most cases when a 0.1 class SM is used. The best results are obtained for cases B and D, using a voltage-based technique. Case B has the highest number of customers and case D mainly consists of small and medium enterprises. It was expected from such feeders that they show high voltage variation and therefore should be easier to allocate. The power-based Pearson correlation method showed worse results than the voltage-based methods. In contrast to the voltage-based techniques, the method was not able to obtain a 100\% accuracy in any case when using a conventional non-`s' SM class when using 20 days of real smart meter data \one{(see Figure \ref{fig:method_comparison})}. 

Ensemble methods can improve performance by leveraging the fact that some consumers show more variation in their voltage while some have more distinctive power variations. Furthermore, ensemble methods can be used to combine different phase identification algorithms and/or different types of data, including time-series measurements collected in different campaigns/points in time.

In a scenario in which power and voltage data are available in equal quality and quantity, the voltage-based method outperforms the power-based method in all but one case. In such a scenario, the DSO would be recommended to use the voltage-based approach. If there is missing voltage data, a bagging ensemble can be used to identify the phase of the remaining customers using their power data. The boosting approach only outperforms the simple bagging ensemble if voltage-based methods obtain a lower accuracy then a power-based method, but performs worse when using low class smart meters.

If supplementary voltage data is only available for a fraction of the consumers in the network, a bagging ensemble improves the accuracy compared to only using the available voltage data, especially when a low accuracy is obtained with the power-based method \one{(see Figure \ref{fig:8})}. The boosting ensemble only outperforms the bagging ensemble in feeders where the power-based method outperforms the voltage-based technique such that the boosting ensemble can make use of the result of the power-based method. This is the case in the studied network where the feeder in case C is better allocated with a power-based method (when a 0.5s class SM is used).

In a scenario with a high roll-out of voltage-measurement capable smart meters, a DSO is recommended to collect supplementary voltage data to use a bagging ensemble only if he is confident that the power data could lead to a decent phase identification. If not, it is best to use a bagging or voltage-based method.

The boosting ensemble often allocates a customer using a power-based method instead of a voltage-based method when the correlation threshold is not well tuned. This shows the main challenge in combining power and voltage-based techniques: it is not clear what accuracy the DSO can expect given a certain scenario. Future work should address techniques to better understand which factors influence the success of ensemble methods, allowing to draw better decision-support guidelines.

\section{Conclusion}
\label{sec:conclusion}
This paper investigated the performance of different state-of-the-art phase identification methods and proposes an ensemble method. \one{To the authors' knowledge, it is the first ensemble method for phase identification that can combine power and voltage data sets. This allows to leverage the advantages of both power- and voltage-based machine learning methods. Furthermore, the two data sets do not need to be collected synchronously, but can be shifted in time.
.}

\one{Our results show that for all accuracy classes the voltage Pearson correlation methods performed better than K-means clustering.} \one{For correlation methods, using transformer measurements as a reference leads to better results  than using a three-phase customer.} For all but one accuracy class, the power-based method is worse than the voltage-based method, for all analyzed feeders. However, voltage measurements are not always available, whereas power-based methods are more accessible, as the same data is used for billing purposes. In general, feeders with fewer customers obtain higher accuracy than feeders with many customers when using the power-based method. It is hard to predict the accuracy of the power-based method a priori, because this varies considerably depending on the type of feeder and the power profiles. However, in general, if the SM accuracy class is high, the accuracy of the power method seems acceptable.

In the future, additional ensemble methods could be proposed, which use different algorithms from the literature and/or different data preprocessing. Furthermore, the factors that influence the success of the different phase identification methods could be further investigated. This would also help to derive an improved metric or set of rules to assess a priori which phase identification methods best suit different scenarios. 

%





\ifCLASSOPTIONcaptionsoff
  \newpage
\fi



%

\bibliographystyle{IEEEtran}
\bibliography{references.bib}

\end{document}